\documentclass[11pt,a4paper]{scrartcl}

\usepackage{CLICdp}

\usepackage{CLICdp_definitions}

\title{All-hadronic HHZ production at 3 TeV CLIC}

\clicdpnote{2020}{003}  

\date{\formatdate{11}{8}{2020}}

\addauthor{Matthias~Weber}{\institute{1}\hcomma\institute{2}}

\addinstitute{1}{CERN, Geneva, Switzerland}
\addinstitute{2}{University of Glasgow, United Kingdom}

\onbehalfof{the CLICdp Collaboration} 

\abstract{In this note, $\PZ\PH\PH$ production in the all-hadronic final state is studied in \epem collisions at the Compact Linear Collider at the \SI{3}{TeV} stage. At this stage this Higgs boson pair production mode is sub-leading to the \ww fusion production cross-section of $\epem\rightarrow\PH\PH\nu\nu$. The events are characterised by a topology of six jets, where the masses of the three pair-wise combinations of two jets are compatible with originating from two \PH and one \PZ bosons. The event selection concentrates on the dominant \PH boson decays into two b-quarks by requiring a presence of multiple b-jets. The study is based on full simulation using the CLICdet model, including beam-induced backgrounds from \gghadrons. Results on the measurement of the total \zhhsm cross section are given.}

\titlecomment{This work was carried out in the framework of the CLICdp Collaboration} 

\graphicspath{ {./logos/}{./figures/} }

\begin{document}

\titlepage

\section{Introduction}
\label{sec:intro}

The Compact Linear Collider (CLIC) is a proposed option for a future electron-positron collider~\cite{Aicheler:2019dhf}. The physics program will be performed in three stages~\cite{CLIC:2016zwp}, running at nominal centre-of-mass energies between \SI{380}{GeV} and \SI{3}{TeV}. The CLIC Higgs physics programme has been discussed in detail in~\cite{Abramowicz:2016zbo}. The two energy stages of \SI{1.5}{TeV} and \SI{3}{TeV} give access to double Higgs boson production measurements, including the extraction of the Higgs self coupling. The study of double Higgs boson production is sensitive to new physics. This can be illustrated using the effective field theory (EFT) formalism. Double Higgs boson measurements at CLIC are discussed in detail in \cite{Roloff:2019crr}. While $\Pep\Pem\rightarrow\zhhsm$ is the dominant double Higgs boson production mode at \SI{1.5}{TeV}, at \SI{3}{TeV} the WW-fusion production mode is dominant, while the \zhhsm production is considerably lower. As shown in \cite{Roloff:2019crr,deBlas:2018mhx}, in the extraction of EFT parameters additional information from \zhhsm is helpful to differentiate between different combinations of parameters. While WW fusion double Higgs boson results have been based on full simulation using the detector model CLIC\_ILD, no \SI{3}{TeV} \zhhsm estimations exist using detector simulation.

In this study the \zhhsm process is investigated for the \SI{3}{TeV} energy stage using full simulation and the new CLICdet detector model. This detector has been developed and studied in several optimisation studies~\cite{CLICdet_note_2017,Arominski:2018uuz} and has been used in a previous analysis of full hadronic \zhsm production at \SI{3}{TeV}~\cite{Leogrande:2019dzm}. The study uses the updated luminosity numbers and the baseline scenario for luminosity sharing between two electron beam polarisation states $(\mathrm{P}(\Pem)=\pm80\%)$ from Ref.~\cite{Robson:2018zje}. 

\section{Detector model and software chain}
\label{sec:DetectorSoftware}

This study is based on the new detector model CLICdet, which is designed to cope with experimental conditions at \SI{3}{TeV} CLIC. 
The superconducting solenoid with an internal diameter of \SI{7}{m} in the centre of the detector provides a magnetic field of \SI{4}{T}. Silicon pixel and strip trackers, the electromagnetic (ECAL) and hadronic calorimeters (HCAL) are situated within the solenoid. Each sub-detector consists of a barrel and two endcap parts. ECAL is a highly granular array of 40 layers of silicon sensors and tungsten plates. HCAL is made out of 60 layers of plastic scintillator tiles read out by silicon photomultipliers, and steel absorber plates. The solenoid is surrounded by the muon system consisting in the endcap of 6, in the barrel of 7 layers of resistive plate chambers interleaved with yoke steel plates. The very forward region of CLICdet is equipped on either side of the interaction point with two smaller electromagnetic calorimeters, LumiCal and BeamCal.

CLICdet uses a right-handed coordinate system, with the origin at the nominal point of interaction. The $z$-axis is along the beam direction, with the electron pointing in the positive direction. The $y$-axis points upwards along the vertical direction. The crossing angle between the electron and positron beams is \SI{20}{mrad}, with $p_{x}^{-}>0$ and $p_{x}^{+}>0$. The polar angle $\theta$ is measured from the positive $z$-axis.

A new software chain for simulation and reconstruction has been developed, based on the DD4hep detector description toolkit~\cite{Frank:2015ivo,Sailer:2017rnh}. The \geant 10.02.p02 toolkit~\cite{Agostinelli:2002hh} is used to simulate the detector response. 
Beam-induced backgrounds from \gghadrons are simulated with \pythia6~\cite{Sjostrand:2006za} using the photon spectra from \guineapig ~\cite{Schulte:382453} as input and CLIC beam parameters at \SI{3}{TeV}. These background collisions are overlaid on the hard physics event. Tracks are reconstructed using the conformal tracking pattern recognition technique~\cite{Brondolin:2019awm}. Software compensation is applied to hits in HCAL to improve the energy measurement, using local energy density information~\cite{Tran:2017tgrSoftwareCompensation}. Pandora particle flow algorithms~\cite{Marshall:2015rfa,Marshall:2012ry} combine information from tracks, calorimeter clusters and muon hits for particle identification and reconstruction. Jet clustering and the jet resolution thresholds ($y_{23}$, $y_{34}$, etc) are calculated using the FastJet 3.3.2 library~\cite{Cacciari:2011ma} . The performance of track reconstruction, particle identification, and flavour tagging at CLICdet has been studied with the new software chain in~\cite{Arominski:2018uuz}. Relative light flavour jet energy resolution values at \SI{3}{TeV} CLIC are typically around 6--8\% for jet energies around \SI{50}{GeV}, decreasing to 4.5--6\% for jet energies larger than \SI{100}{GeV}, and about 3-4\% for \SI{1}{TeV} jets.

\section{HHZ signal reconstruction}
\label{sec:HHZSignal}

\begin{figure}[htbp!]
\centering
\begin{minipage}[l]{0.32\textwidth}
\includegraphics[width=1.0\textwidth]{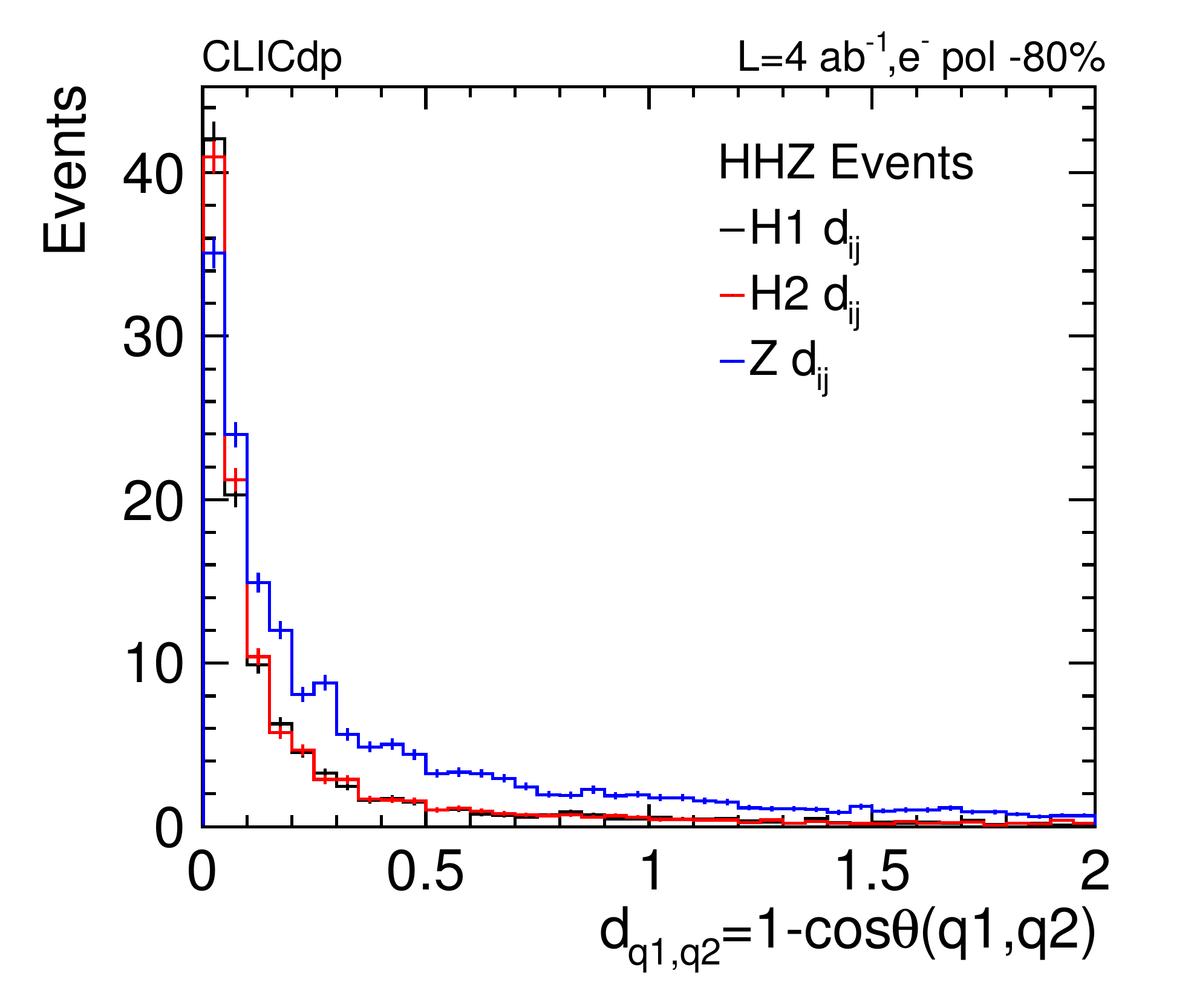}
\end{minipage}
\begin{minipage}[r]{0.32\textwidth}
\includegraphics[width=1.0\textwidth]{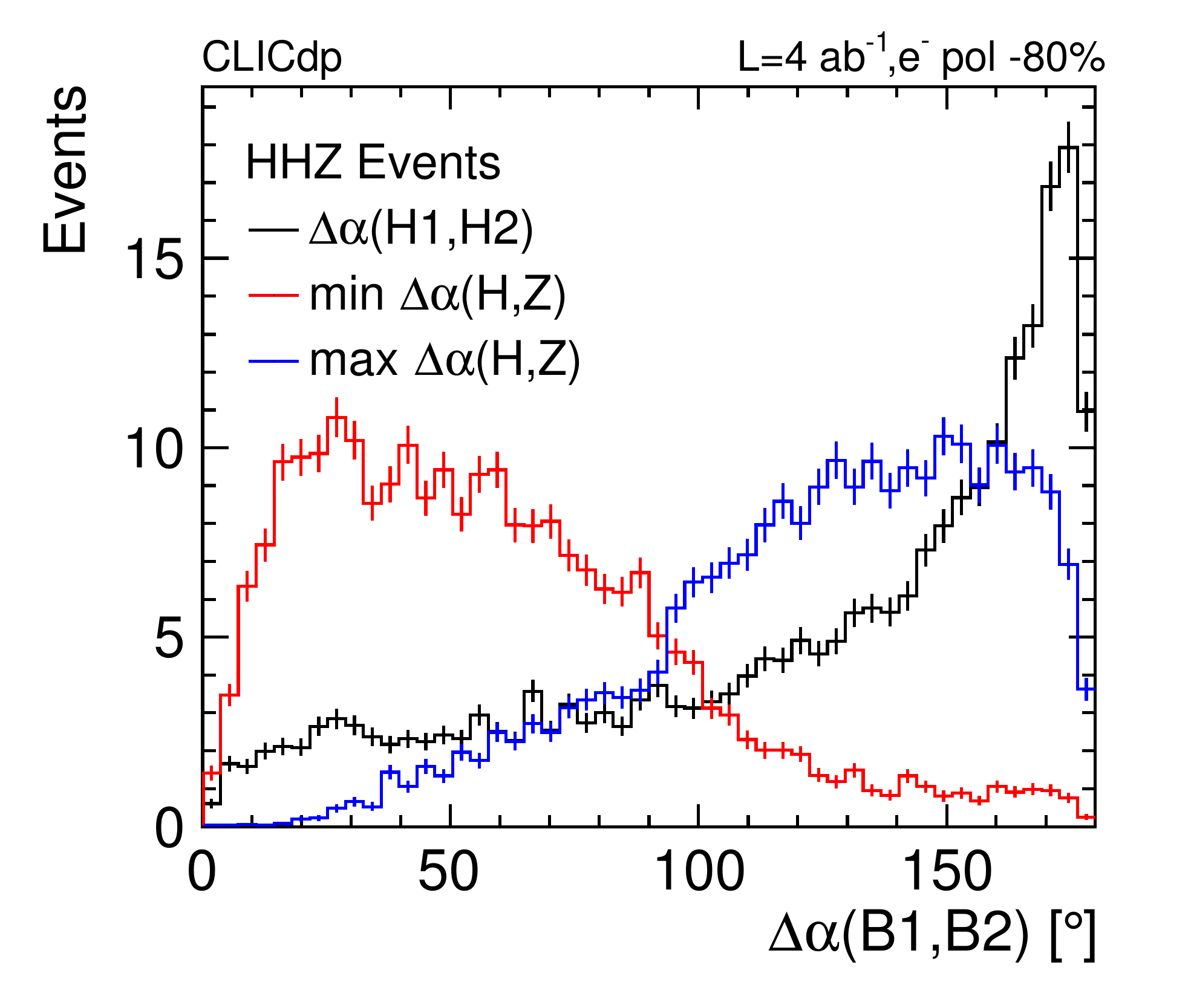}
\end{minipage}
\begin{minipage}[l]{0.32\textwidth}
\includegraphics[width=1.0\textwidth]{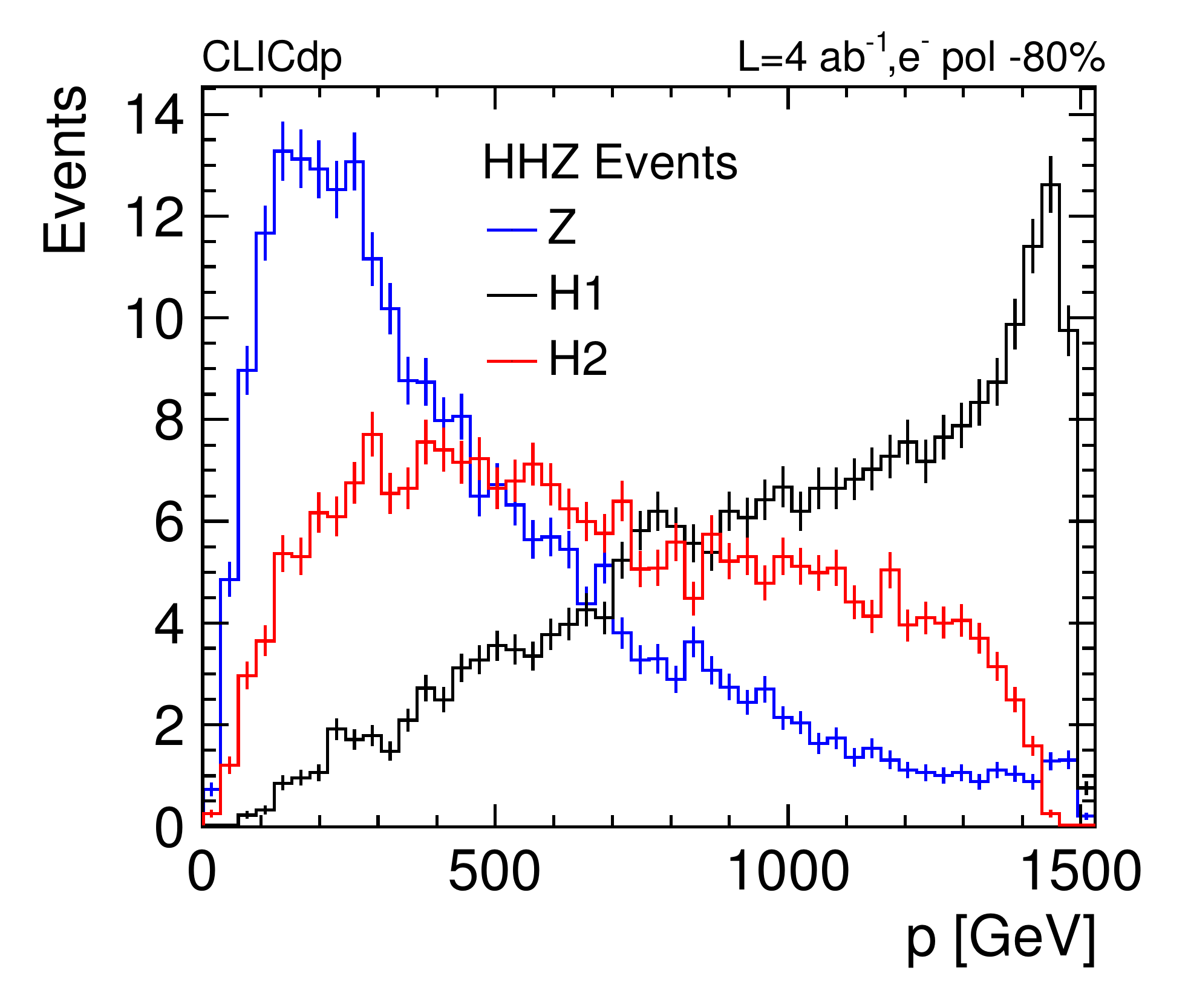}
\end{minipage}
\caption{The distance measure between the two quarks of the boson decay in all hadronic events with $\PZ\PH\PH\rightarrow \qqbar\bb\bb$ (left), the angles between the three bosons for all $\PZ\PH\PH$ events (middle), and the momenta $p$ of the three bosons for all $\PZ\PH\PH$ events (right).}
\label{fig:B_dij_B_qq_dalpha}
\end{figure}

In this study we consider all hadronic decays of the \PZ boson, while for the \PH boson the dominant Standard Model decay mode $\PH\rightarrow\bb$ with a branching ratio of around 58.4\% is considered. Thus in the following the signal phase-space is about $24\%$ of the total $\PZ\PH\PH$ production.
Typically the distance measure $d_{q1,q2}=1-\cos\theta(q1,q2)$ is larger for the hadronic \PZ decays than the one from the \PH boson decays considering all hadronic decays with $\PH\PH\rightarrow \qqbar\bb\bb$ (Fig.~\ref{fig:B_dij_B_qq_dalpha} left). Here the angle $\theta$ refers to the opening angle between the momenta of the two quarks $q1$ and $q2$. The opening angle $\Delta\alpha$ between both \PH bosons is larger than the angles between the \PZ and any of the \PH bosons as shown in the Fig.~\ref{fig:B_dij_B_qq_dalpha} middle. On average the \PZ boson is the least boosted, while the \PH boson with the larger momentum carries typically the largest momentum of all bosons as displayed in Fig.~\ref{fig:B_dij_B_qq_dalpha} right.

\begin{figure}[htbp!]
\centering
\begin{minipage}[l]{0.49\textwidth}
\includegraphics[width=1.0\textwidth]{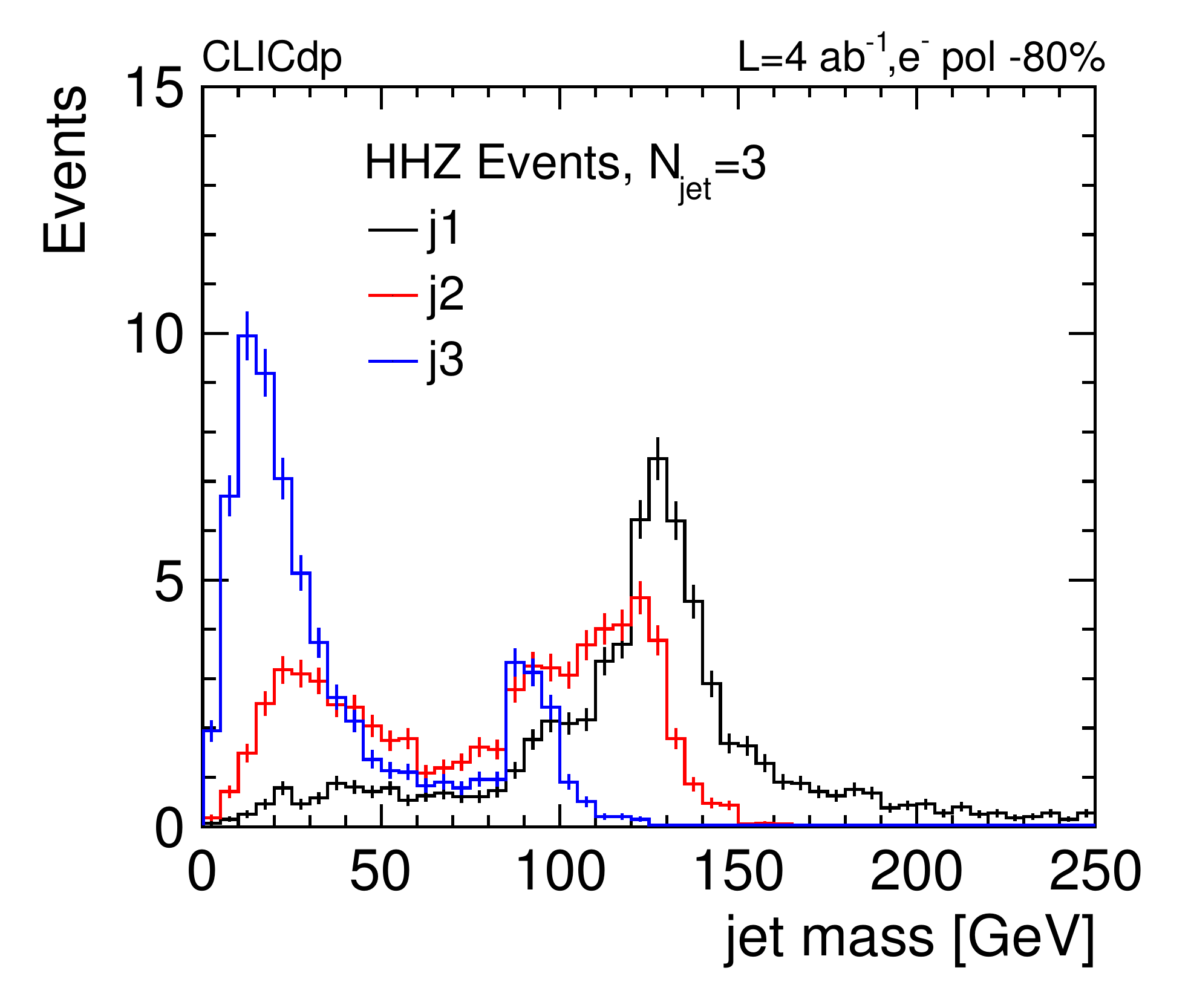}
\end{minipage}
\begin{minipage}[r]{0.49\textwidth}
\includegraphics[width=1.0\textwidth]{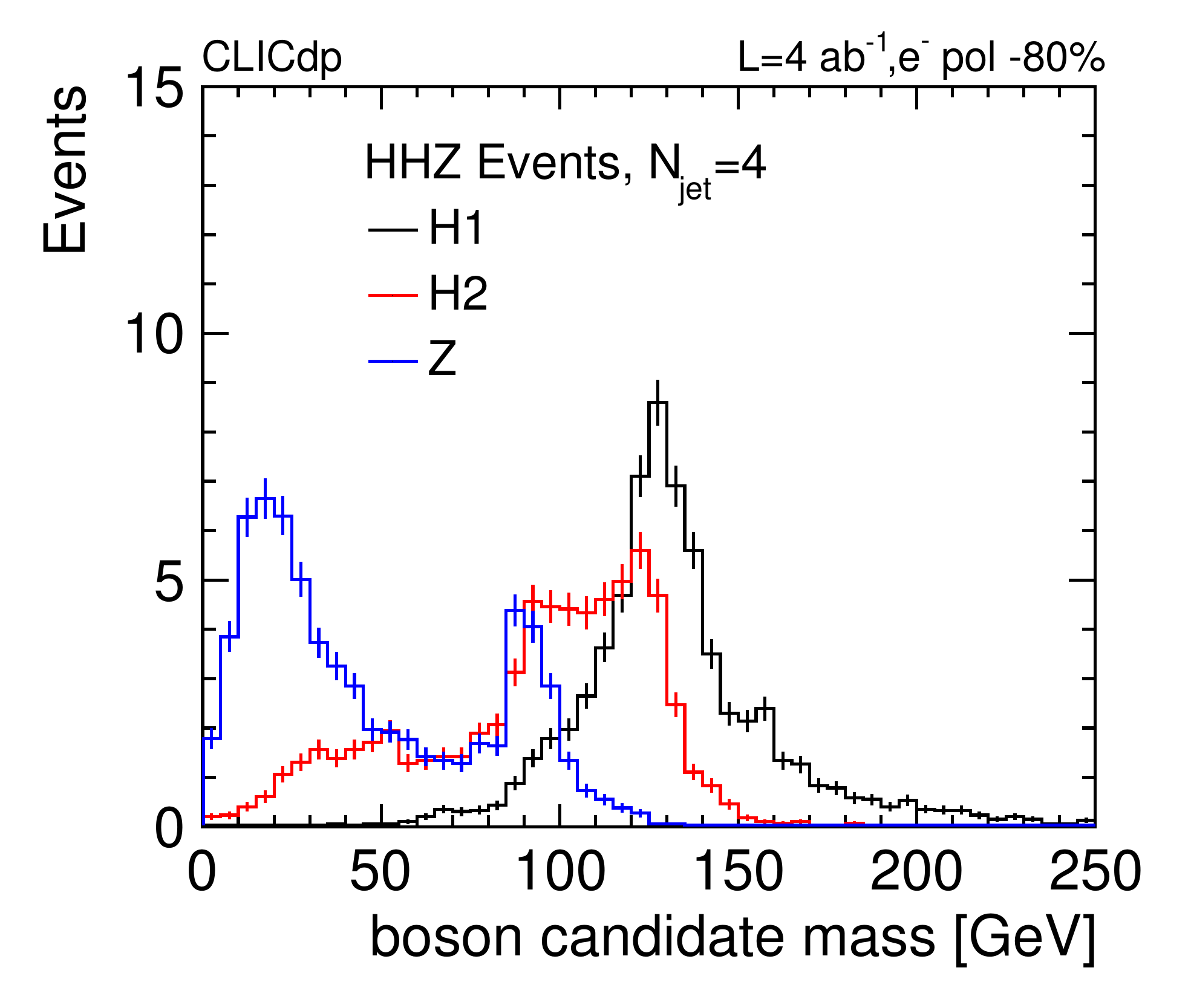}
\end{minipage}
\begin{minipage}[l]{0.49\textwidth}
\includegraphics[width=1.0\textwidth]{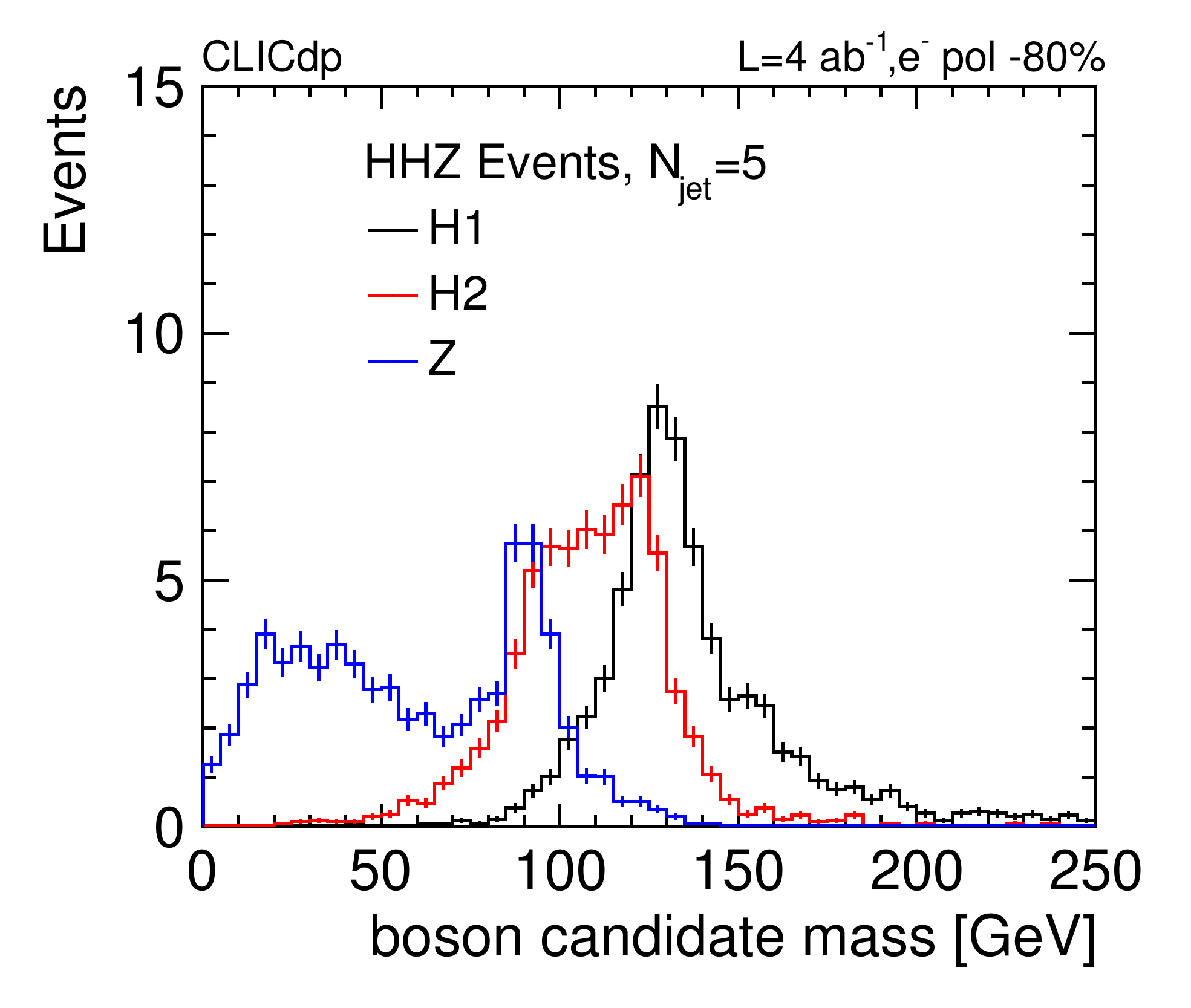}
\end{minipage}
\begin{minipage}[r]{0.49\textwidth}
\includegraphics[width=1.0\textwidth]{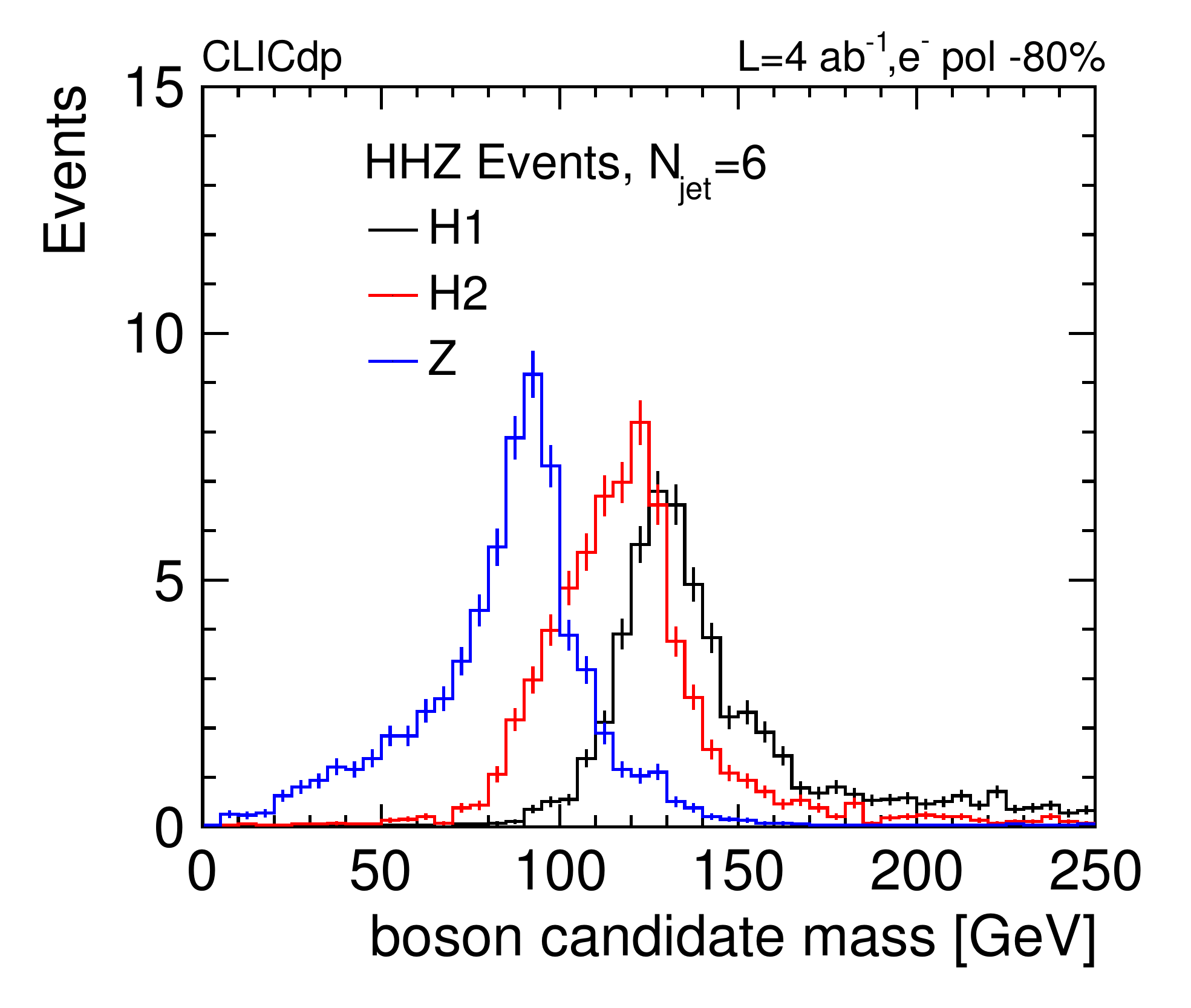}
\end{minipage}
\caption{Masses of three boson candidates for for all hadronic $\PZ\PH\PH$ decays with $\PH\PH\rightarrow\bb\bb$ in events using exclusive jet clustering into three (top left), four (top right), five (bottom left), and six (bottom right) jets, using the VLC algorithm with $\beta=\gamma=1$ and a radius $R=1.1$.}
\label{Fig:HHZ_comb_3_rj}
\end{figure}

Inspired by a study of boosted hadronic HZ events, in a first attempt events are reconstructed using three large cone jets. These jets are reconstructed with the VLC algorithm~\cite{Boronat:2016tgd} as implemented in the FastJet library~\cite{Cacciari:2011ma} with a radius $R=1.1$ and $\gamma=\beta=1$ in exclusive mode to force the event into three jets. This large cone was chosen in order to collect most of the event energy of all $\PZ\PH\PH$ topologies. The partonic \roots agrees with the \roots calculated from the three MC truth particle level jets after adding the four vector of the neutrinos, which are present in decays of B hadrons, confirming that the majority of the visible event energy is collected in these three jets. 
 While on detector level the total visible event energy is well reconstructed using three detector level jets, the mass reconstruction is not that satisfactory (see top left Fig.~\ref{Fig:HHZ_comb_3_rj}). Clustering the event into three jets shows only a clear mass peak for the jet with the largest jet mass, the jet with the lowest mass shows a peak around the \PZ-mass, but many more events appear at a second lower mass peak. A clear two-peak structure is observed for the jet with the second highest jet mass. The second peak at higher masses is spread spanning both the \PZ-mass and the \PH-mass. Jet clustering into three jets is not adequate for most events, particularly for the jet closest to the \PZ boson direction, which is on average the boson with smallest momentum. The exclusive jet clustering is tested in three more configurations forcing the event into four, five, and six jets. In events with four jet clustering two jets are combined into one \PH/\PZ boson candidate; in events where five jet clustering is studied, four jets are pairwise combined into two \PH/\PZ boson candidates; and in six jet clustering all six jets are pairwise combined into three final \PH/\PZ boson candidates. The combination is chosen by minimising the sum

\begin{equation}
\mathrm{sum}=\min\sum_{\mathrm{comb}}((m(H1)-m_{\mathrm{H}})^{2}+(m(H2)-m_{\mathrm{H}})^{2}+(m(Z)-m_{\mathrm{Z_{0}}})^{2}),
\label{eq:sumMasses}
\end{equation}
where $H1$, $H2$, and $Z$ are the combined \PH and \PZ boson candidates, and $m_{\mathrm{H}}$ and $m_{\mathrm{Z_{0}}}$ are the \PH and \PZ boson masses. The best mass reconstruction is achieved by clustering the event into six jets (see Fig.~\ref{Fig:HHZ_comb_3_rj}). Increasing or decreasing the VLC jet radius in the jet clustering to $R=1.5$ or $R=0.7$ respectively leads to similar results. Using three \PZ/\PH boson candidates from four or five original jets, two clear peaks appear for the \PH candidates, but the \PZ boson candidate mass is still spread over two peaks. Only in the six jet configuration after pairwise combining these in three final boson candidates three clear mass peaks appear with the lowest peak close to the \PZ-mass and two higher peaks close to the \PH-mass, at the expense of a tail to larger masses for the combination with the largest mass. The width of the peaks is smallest using six jets as well. Therefore the VLC algorithm with $R=1.1$, $\gamma=\beta=1$, and $\mathrm{N}=6$ is used in the exclusive jet clustering for the following analysis.

\begin{figure}[htbp!]
\centering
\begin{minipage}[l]{0.49\textwidth}
\includegraphics[width=1.0\textwidth]{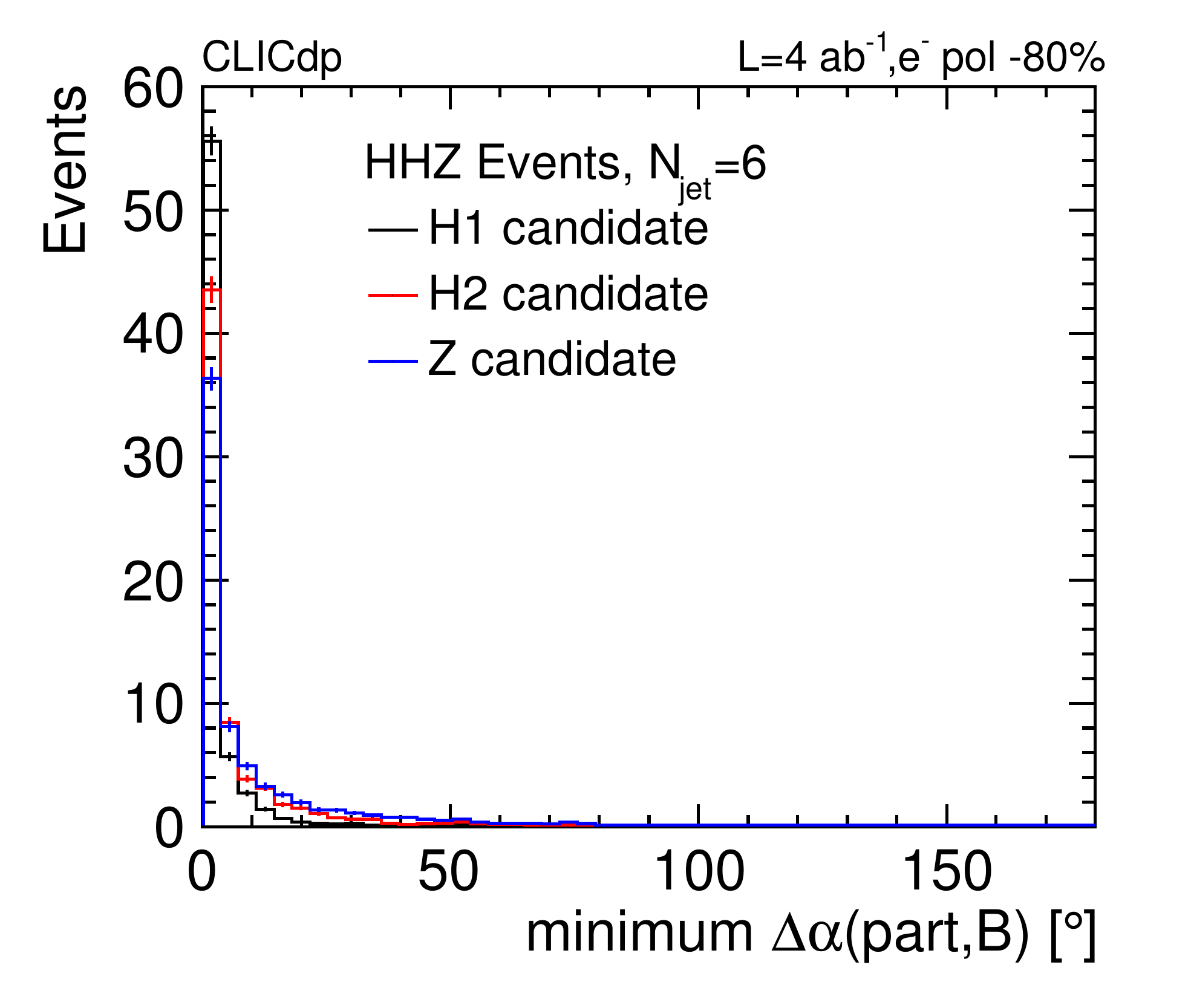}
\end{minipage}
\begin{minipage}[l]{0.49\textwidth}
\includegraphics[width=1.0\textwidth]{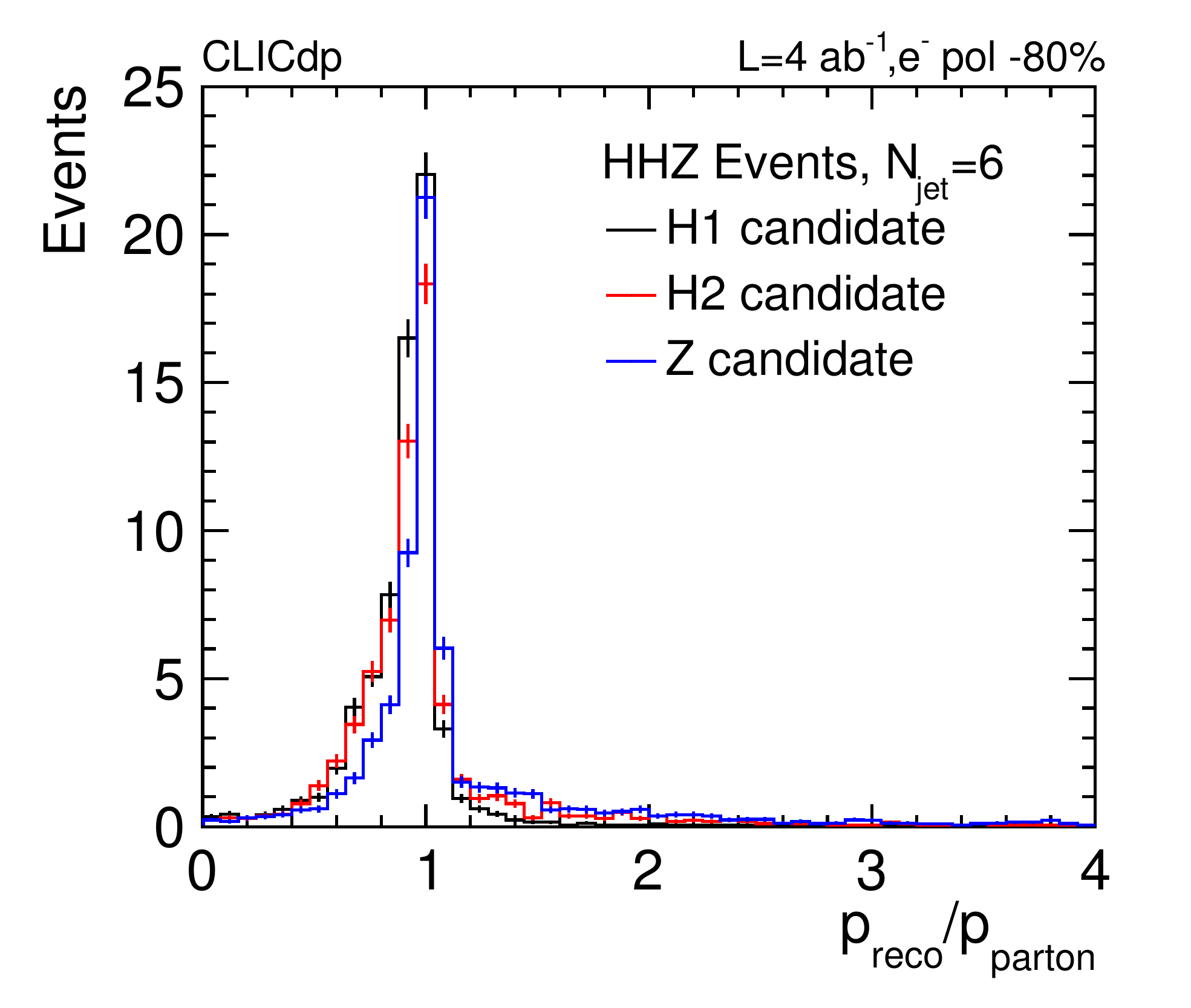}
\end{minipage}
\caption{The angles between the boson candidates and the closest boson on parton level, (left) and the response of the boson candidate momentum and the matched boson parton level momentum (right) both for events with $\PZ\PH\PH\rightarrow \qqbar\bb\bb$.}
\label{Fig:jet_comb_vs_parton_info}
\end{figure}

The resulting three \PH and \PZ boson candidates are compared to the three bosons on parton level in Fig.~\ref{Fig:jet_comb_vs_parton_info} in the phase-space of $\PZ\PH\PH\rightarrow\qqbar\bb\bb$. In general each of the three bosons are close to one of the original partons, particularly for the reconstructed candidate with the largest mass, referred to as H1, which typically corresponds to the object with the largest momentum as well. Comparing the momentum $p_{\mathrm{reco}}$ of the three boson candidates with the momentum of the matched boson on parton level $p_{\mathrm{parton}}$  a clear peak at 1 can be observed for all candidates. The elongated tails of both \PH candidate objects to lower reconstructed momenta reflect the fact that neutrinos in B-hadron decays escape detection. In decays of hadrons in the \PZ boson decay chain neutrinos play a less prominent role, and thus the reconstructed \PZ candidate response is more symmetric. While the combination procedure works well, future improvements could be achieved by modifying the summation in eq.~\ref{eq:sumMasses} to reflect different resolutions for the reconstructed invariant masses for the \PZ and Higgs bosons, taking into account the long tail to lower masses for the \PH candidates as well. The jet clustering and the size of the jet cone impacts the means of the invariant mass distributions, which could be corrected for after a detailed study.

The identification of $b$-jets is performed by the linear collider flavour identification (LCFIPlus) tool~\cite{Suehara:2015ura}. The identification starts with a primary vertex finder, and continues with the identification of secondary vertices to identify $b$ and $c$ hadron decays. The secondary vertices are attached to jets. Isolated leptons within the jets are checked for compatibility with secondary vertices originating from semi-leptonic decays of heavy flavour hadrons. In the refined jet clustering step particles are combined into jets with the VLC jet algorithm using the tracks and leptons originating from secondary vertices as seed. Values are attached to each jet reflecting its compatibility with originating from $b$ (BTag), $c$ (CTag), and light flavour quarks (LTag). It is investigated if the pairwise jet combination can be improved using BTag information, since at least four b-quarks are involved in the decays of signal events. These attempts lead to no improvement of the mass resolution, energy and spatial agreement with the underlying bosons on parton level, thus BTagging information is not used as additional input in the combination.

\section{Monte Carlo simulation}
\label{sec:MCSimulation}

Both signal and backgrounds samples are produced by \whizard2.7.0~\cite{Kilian:2007gr}, using luminosity spectra from \guineapig interfaced by \textsc{circe2}{} with initial state radiation enabled. Parton shower and hadronisation are handled by \pythia6~\cite{Sjostrand:2006za}. The \textsc{DD4hep}{} detector description toolkit has been used to implement the simulated model of CLICdet in \geant{}, version 10.02p02, via the \textsc{DDG4} package. \\
Backgrounds to all-hadronic \zhhsm events originate from di-quark $\epem\rightarrow \qqbar$, four-quark $\epem\rightarrow \qqqq$ and six-quark $\epem\rightarrow \qqqqqq$ final states, as well as triboson backgrounds from ZZH and WWH with hadronically decaying \PZ and \PW bosons. Table~\ref{Tab:MC_samples} lists the details of the produced samples for both negative and positive polarisation of 80\% of the electron beam. The weight of each event is calculated under the assumption of luminosity sharing of the ratio 4:1 between the negative and positive polarisation of the electron beam, thus $\mathrm{L}_{-80\%}=\SI{4}{\abinv}$ and $\mathrm{L}_{+80\%}=\SI{1}{\abinv}$ are used as values for the integrated luminosity. The polarisation has a moderate impact on the $\PZ\PH\PH$ signal, decreasing the cross section by about 45\% for positive compared to negative electron beam polarisation, a similar impact can be observed for the di-quark sample. The four-quark production cross section is dominated by WW boson production and thus largely reduced for positive polarisation by a factor of about 7.5. The six-quark dataset is split into 23 samples to cover all possible flavour combinations compatible with \ttbar and tri-boson production. In table~\ref{Tab:MC_samples}  the six-quark flavour combinations with the largest cross sections are shown. For the six-quark dataset positive polarisation reduces the cross section considerably as well. 

\begin{table}[hbtp]
 \centering
 \caption{Signal and background datasets with $y=\PQd,\PQs,\PQb$, $\mathrm{L}=\SI{4}{\abinv}$ for P(\Pem)=-80\%, $\mathrm{L}=\SI{1}{\abinv}$ for P(\Pem)=+80\%:}
 \begin{tabular}{|c|c|c|c|c|c|}
\hline
process =& Events & $\sigma$[fb] & Polarisation & event weight \\
\hline
$\epem\rightarrow\PH\PH\qqbar$  & 9600 & 4.18e-2 & P(\Pem)=-80\% & 0.0174 \\
$\epem\rightarrow\PH\PH\qqbar$  & 9552 & 2.30e-2 & P(\Pem)=+80\% & 0.00303 \\
$\epem\rightarrow\zhsm$ &  114000 &3.83 & P(\Pem)=-80\% & 0.134 \\
$\epem\rightarrow\zhsm$ &  27840 & 2.76 & P(\Pem)=+80\% & 0.0959 \\
$\epem\rightarrow\qqbar$ &  1549464 & 1269 & P(\Pem)=-80\% & 3.28 \\
$\epem\rightarrow\qqbar$ &  388392 & 786 & P(\Pem)=+80\% & 2.02 \\
$\epem\rightarrow\qqqq$ &  1915464 & 902 & P(\Pem)=-80\% & 1.88 \\
$\epem\rightarrow\qqqq$ &  479040 & 120 & P(\Pem)=+80\% & 0.251 \\
$\epem\rightarrow\PQd\PQd\PQu  yy\PQu$ &  456336 & 14.5 & P(\Pem)=-80\% & 0.127\\
$\epem\rightarrow\PQd\PQd\PQu  yy\PQu$ &   121200 & 5.01 & P(\Pem)=+80\% & 0.0413 \\
$\epem\rightarrow yy\PQu\PQb\PQb\PQc$ & 428405  & 13.3 & P(\Pem)=-80\% & 0.124\\
$\epem\rightarrow yy\PQu\PQb\PQb\PQc$ &  123720 & 5.21 & P(\Pem)=+80\% & 0.0421\\
$\epem\rightarrow\PQs\PQs\PQc\PQb\PQb\PQc$ &  330096 & 12.5 & P(\Pem)=-80\% & 0.151\\
$\epem\rightarrow\PQs\PQs\PQc\PQb\PQb\PQc$ &  84240 & 4.89 & P(\Pem)=+80\% & 0.0581 \\
$\epem\rightarrow\PZ\PZ\PH\rightarrow\qqqq\PH$  & 5784 & 1.39e-01 & P(\Pem)=-80\% & 0.0964\\
$\epem\rightarrow\PZ\PZ\PH\rightarrow\qqqq\PH$   & 2904 & 7.16e-02 & P(\Pem)=+80\% & 0.0247 \\
$\epem\rightarrow\PW\PW\PH\rightarrow\qqqq\PH$  & 94608 & 4.116 & P(\Pem)=-80\% & 0.0174 \\
$\epem\rightarrow\PW\PW\PH\rightarrow\qqqq\PH$   & 2424 & 5.176e-01 & P(\Pem)=+80\% & 0.214 \\
\hline
  \end{tabular}
    \label{Tab:MC_samples} 
 \end{table}

\section{Preselection}
\label{sec:variables}

\begin{figure}[htbp!]
\centering
\begin{minipage}[l]{0.32\textwidth}
\includegraphics[width=1.0\textwidth]{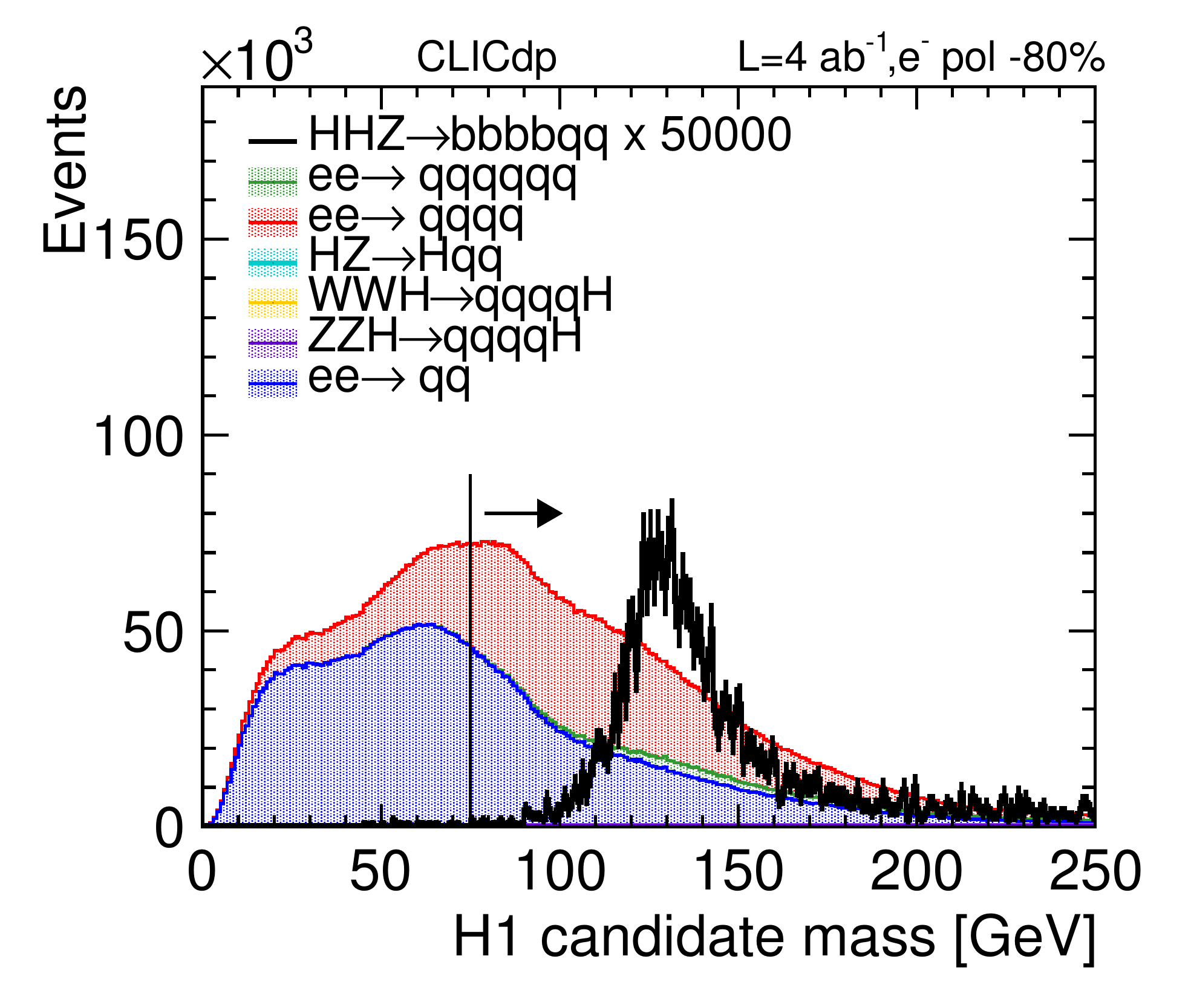}
\end{minipage}
\begin{minipage}[r]{0.32\textwidth}
\includegraphics[width=1.0\textwidth]{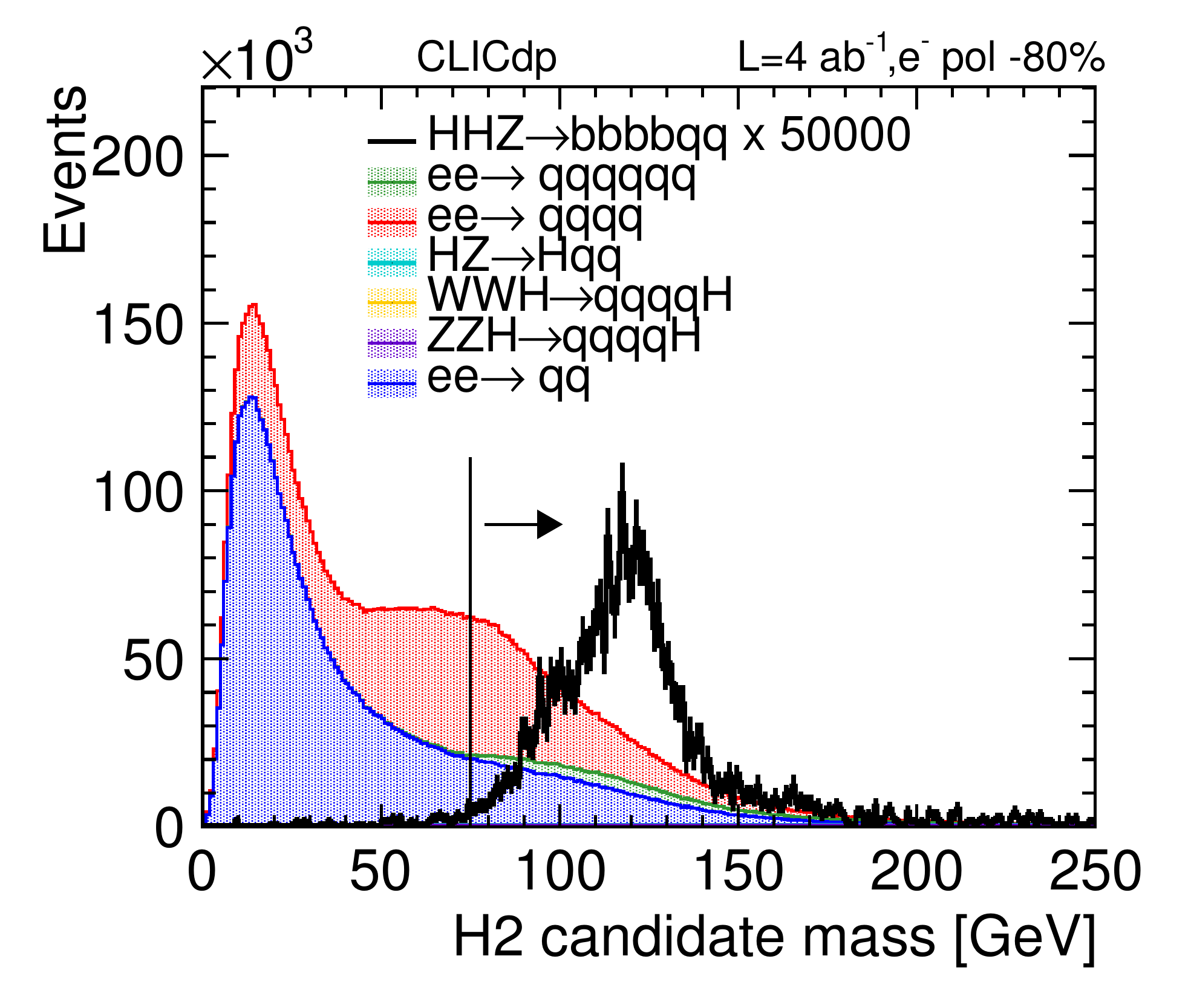}
\end{minipage}
\begin{minipage}[l]{0.32\textwidth}
\includegraphics[width=1.0\textwidth]{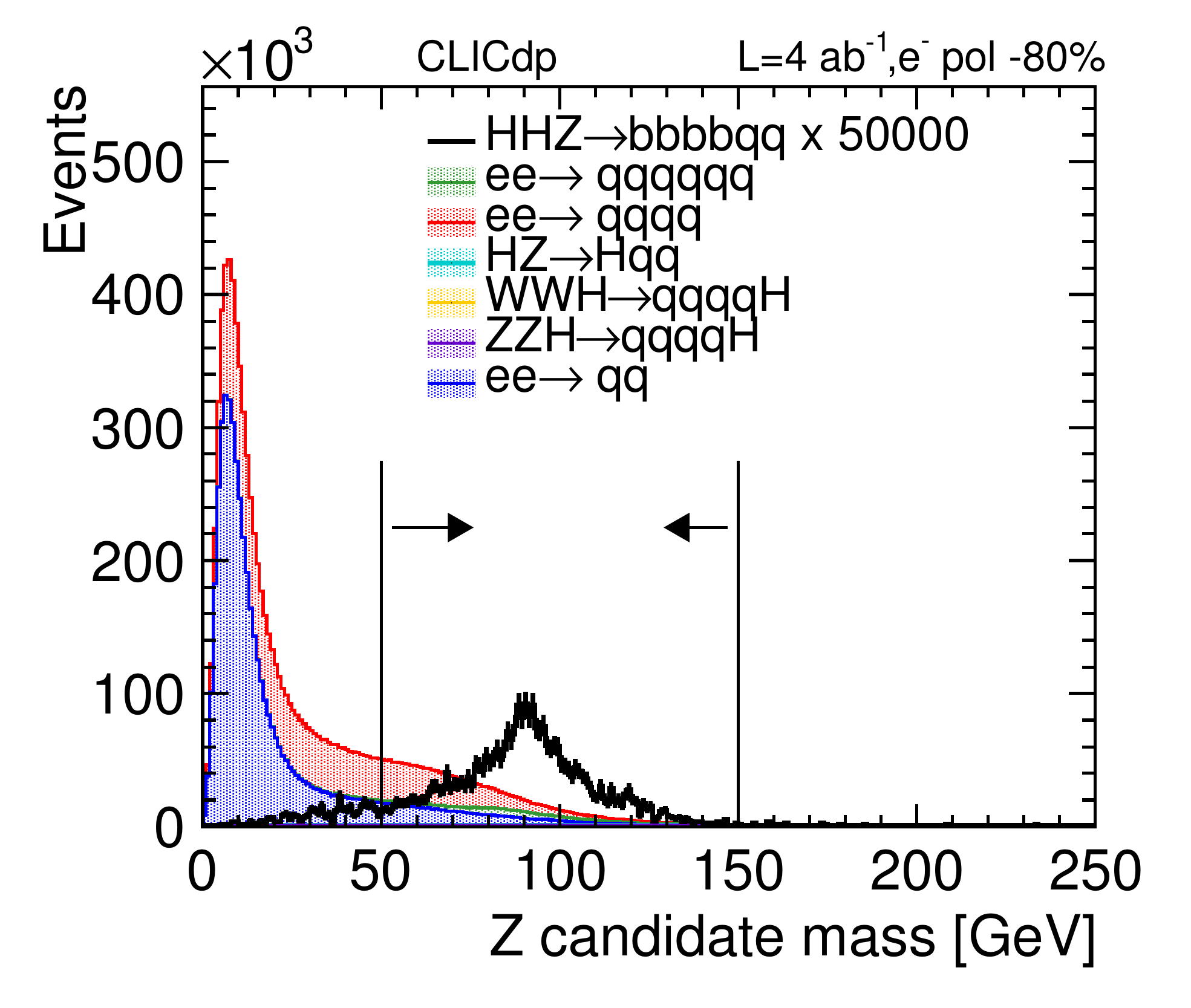}
\end{minipage}
\caption{The mass distributions for the more massive (left) and less massive (middle) \PH boson candidates, and the \PZ boson candidate (right) for all background events from \zhsm, $\epem\rightarrow\qqbar$, $\epem\rightarrow\qqqq$ , and $\epem\rightarrow \text{qqqqqq}$ combined and signal events $\PZ\PH\PH\rightarrow \qqbar\bb\bb$ weighted by a factor of 50 000. The vertical lines and arrows indicate the signal selection.}
\label{fig:bosonMasses}
\end{figure}

\begin{figure}[htbp!]
\centering
\begin{minipage}[l]{0.32\textwidth}
\includegraphics[width=1.0\textwidth]{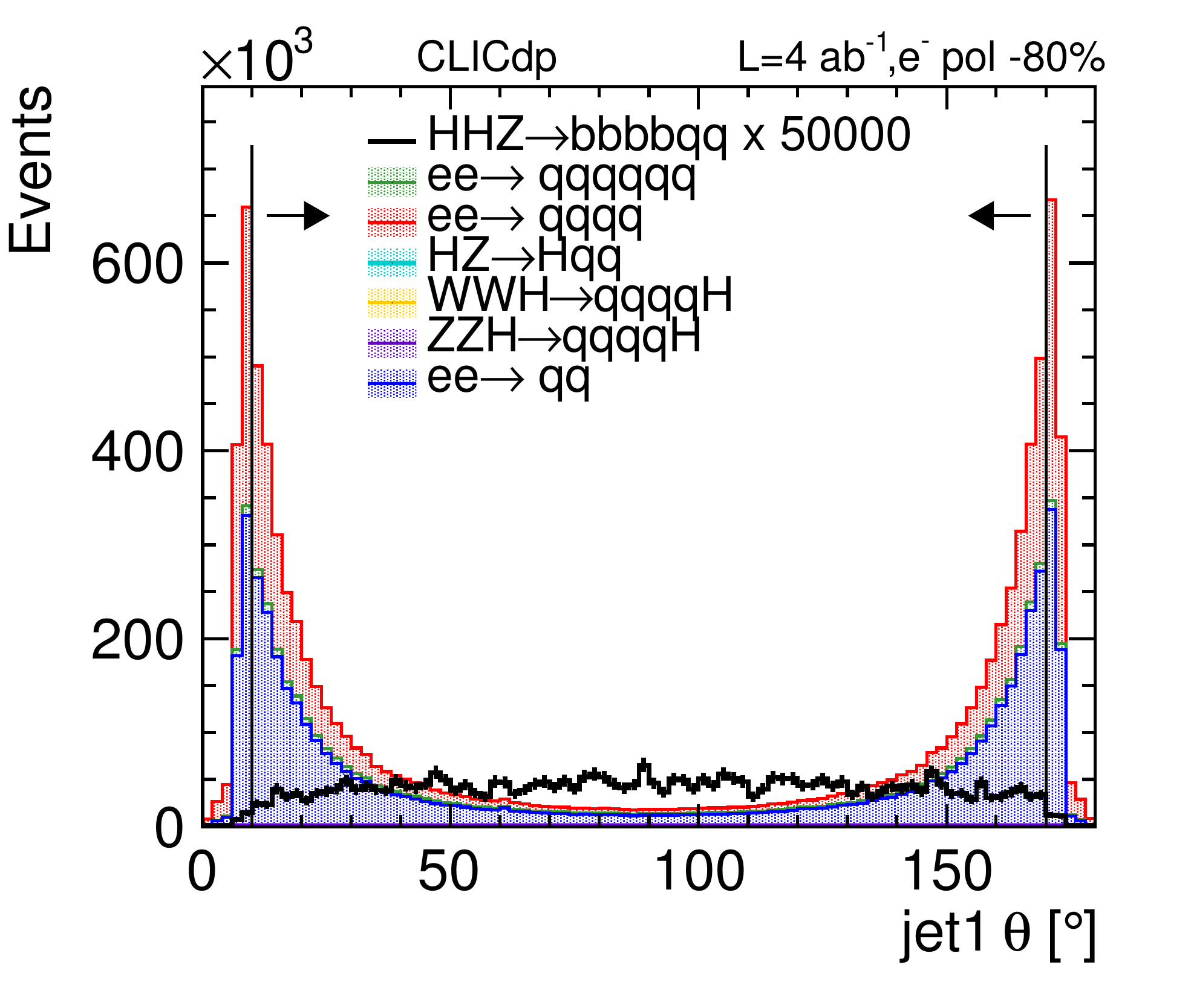}
\end{minipage}
\begin{minipage}[r]{0.32\textwidth}  
\includegraphics[width=1.0\textwidth]{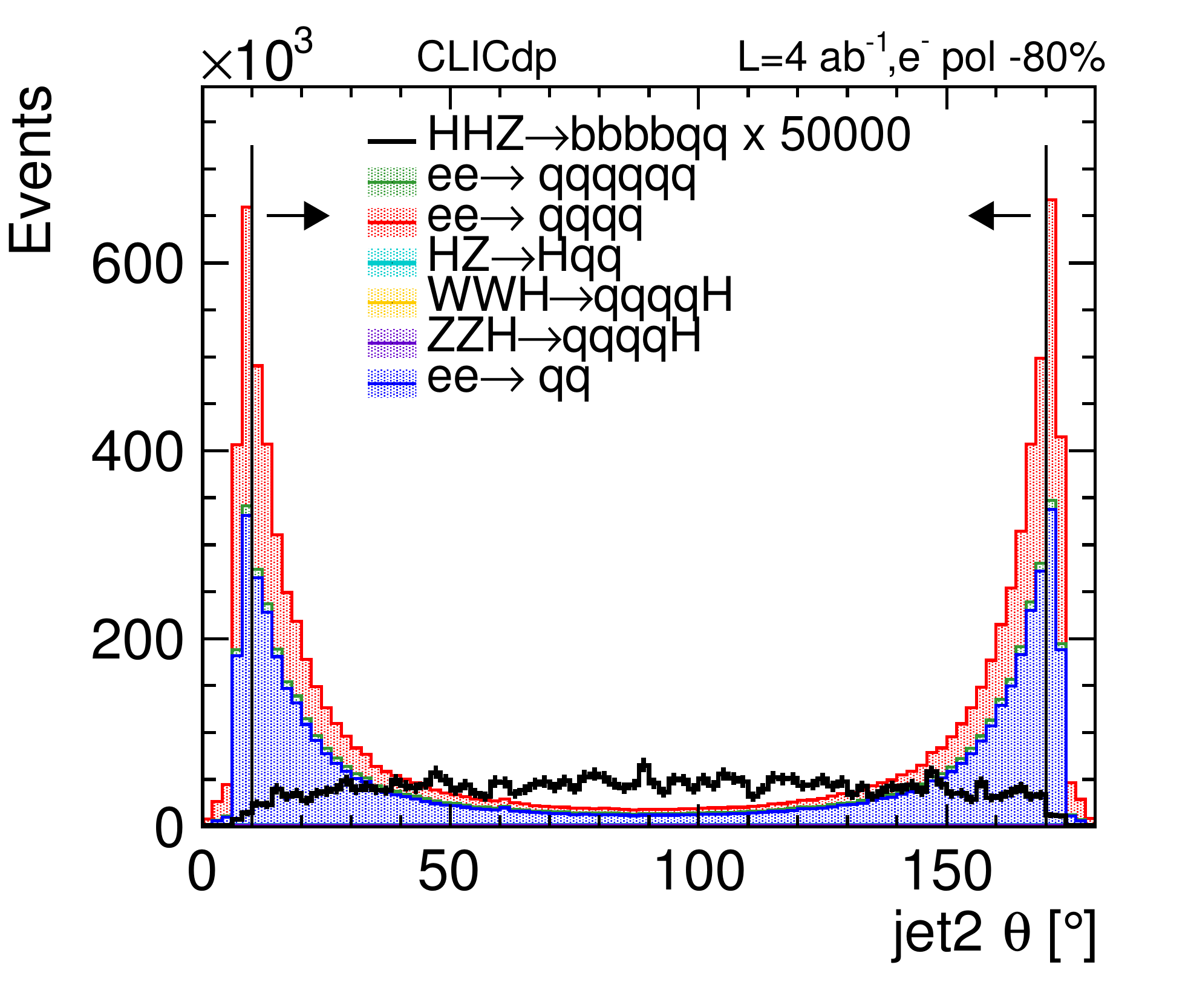}
\end{minipage}
\begin{minipage}[l]{0.32\textwidth}
\includegraphics[width=1.0\textwidth]{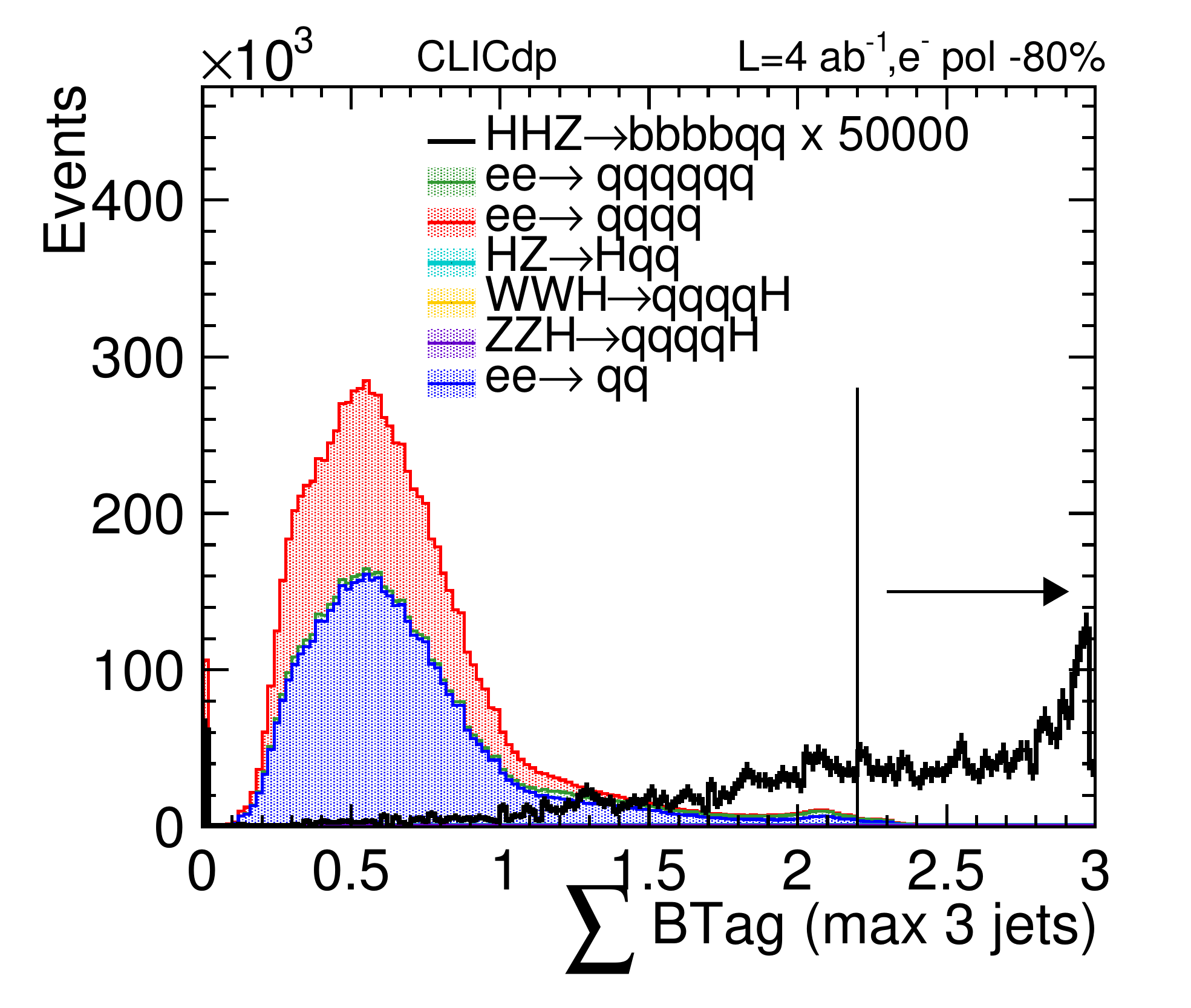}
\end{minipage}
\caption{The polar angle distribution for background events from \zhsm, $\epem\rightarrow\qqbar$, $\epem\rightarrow\qqqq$ , and $\epem\rightarrow \text{qqqqqq}$ combined and signal events $\PZ\PH\PH\rightarrow \qqbar\bb\bb$ weighted by a factor of 50 000, for the jet with largest energy (left) and the jet with second largest energy (middle), as well as the BTag sum of the three jets with the largest BTag values (right). The vertical lines and arrows indicate the signal selection.}
\label{fig:polarangles_BTagSum}
\end{figure}

\begin{figure}[htbp!]
\centering
\begin{minipage}[l]{0.25\textwidth}
\includegraphics[width=1.0\textwidth]{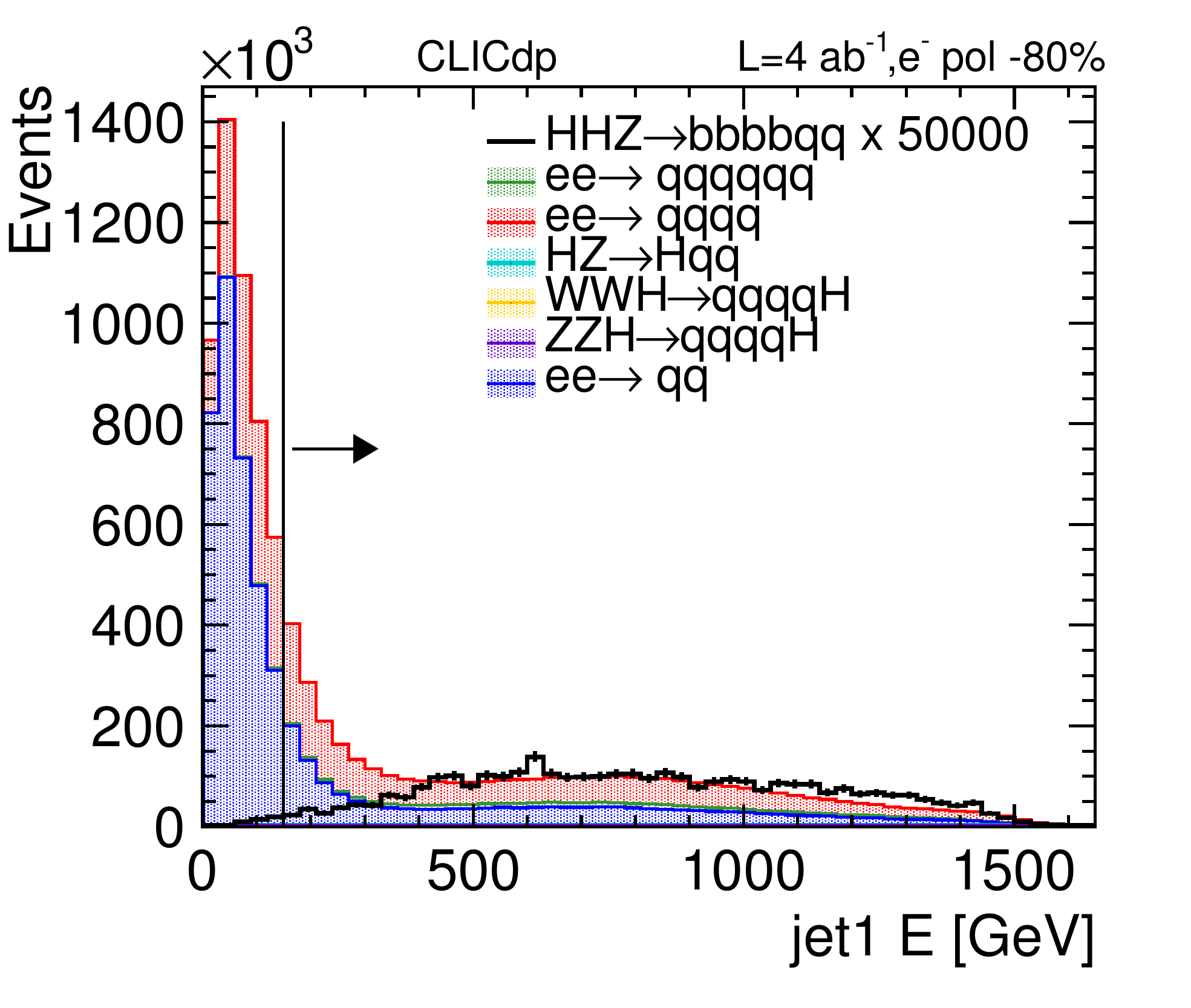}
\end{minipage}%
\begin{minipage}[r]{0.25\textwidth}
\includegraphics[width=1.0\textwidth]{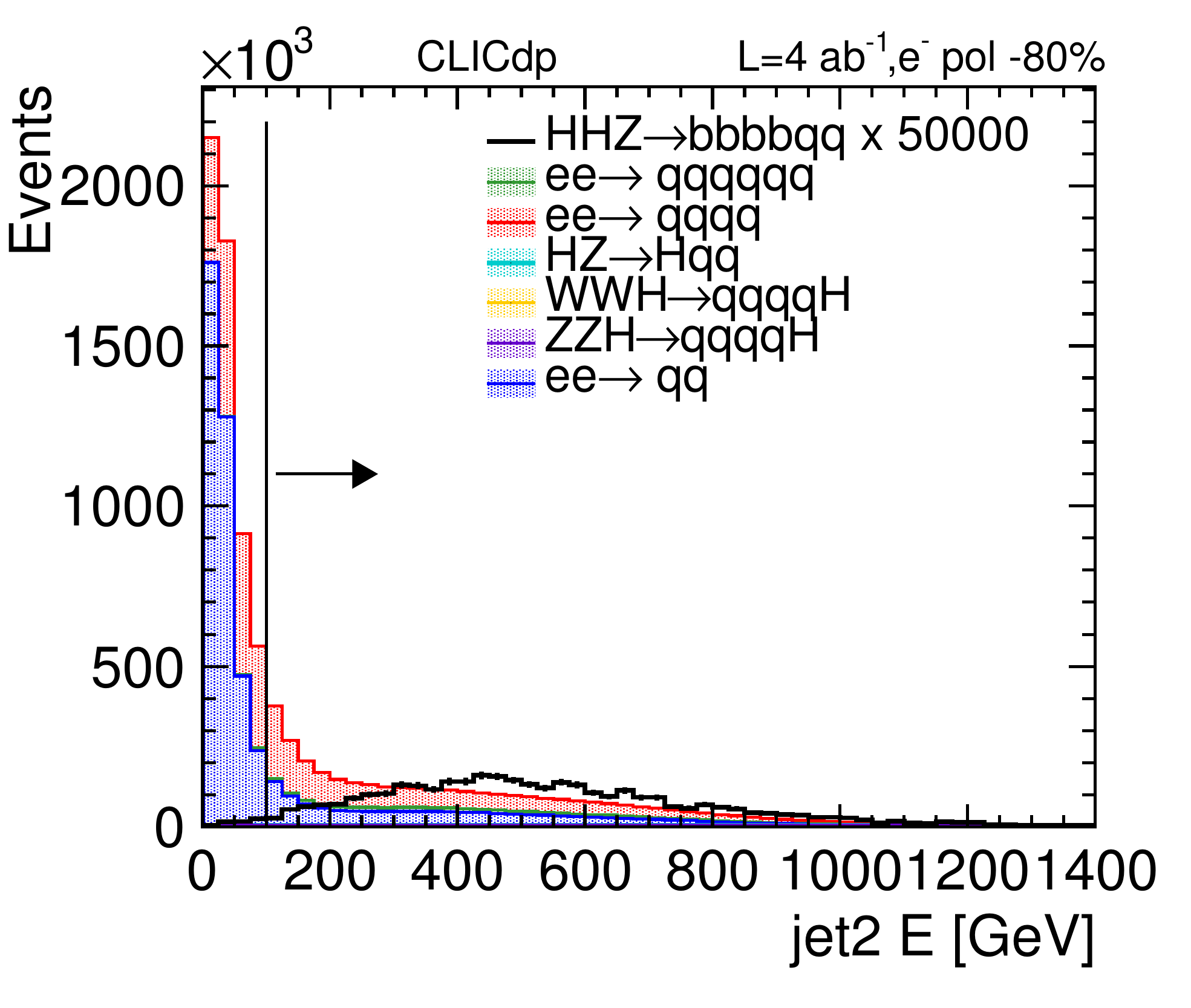}
\end{minipage}%
\begin{minipage}[r]{0.25\textwidth}
\includegraphics[width=1.0\textwidth]{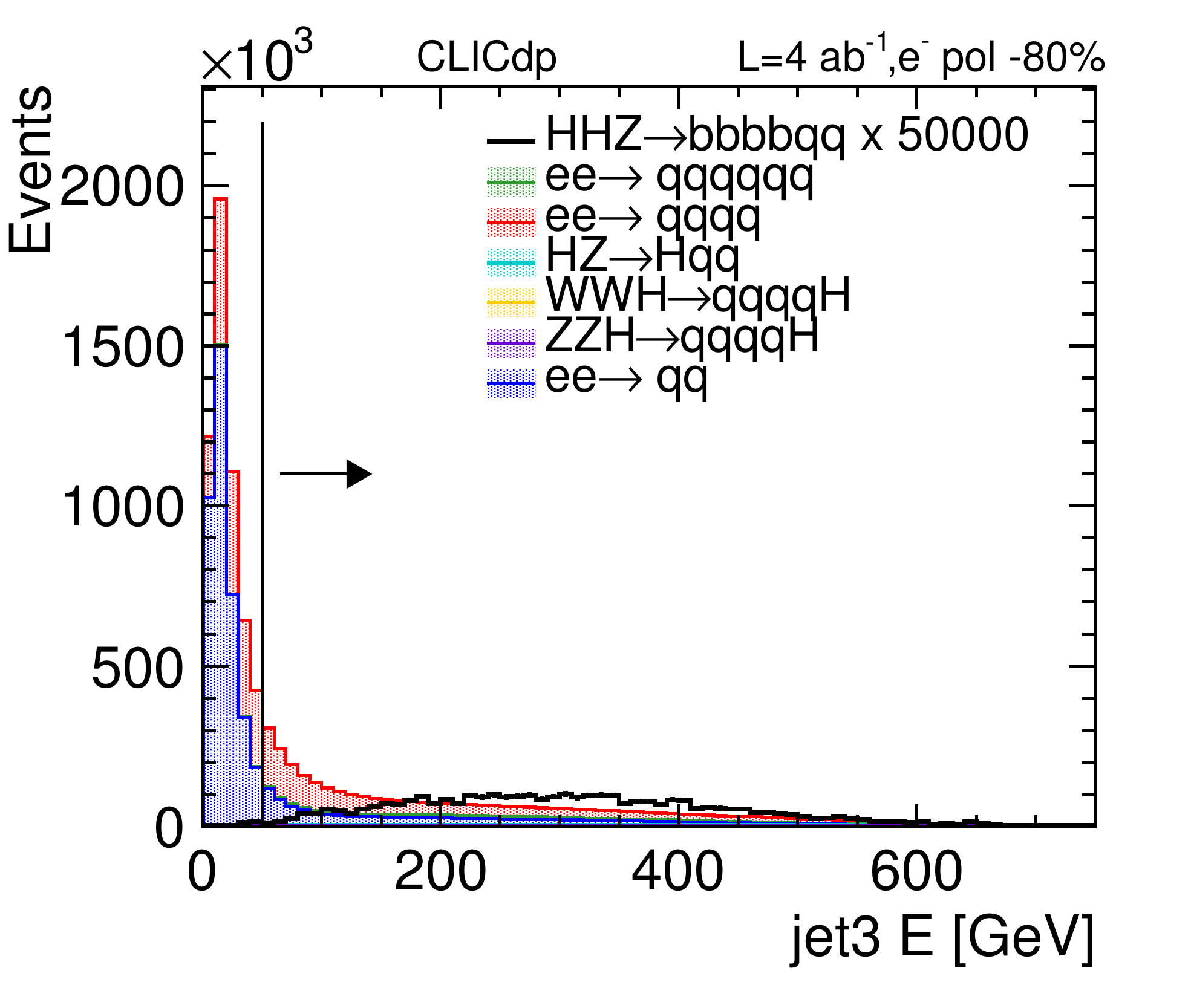}
\end{minipage}%
\begin{minipage}[r]{0.25\textwidth}
\includegraphics[width=1.0\textwidth]{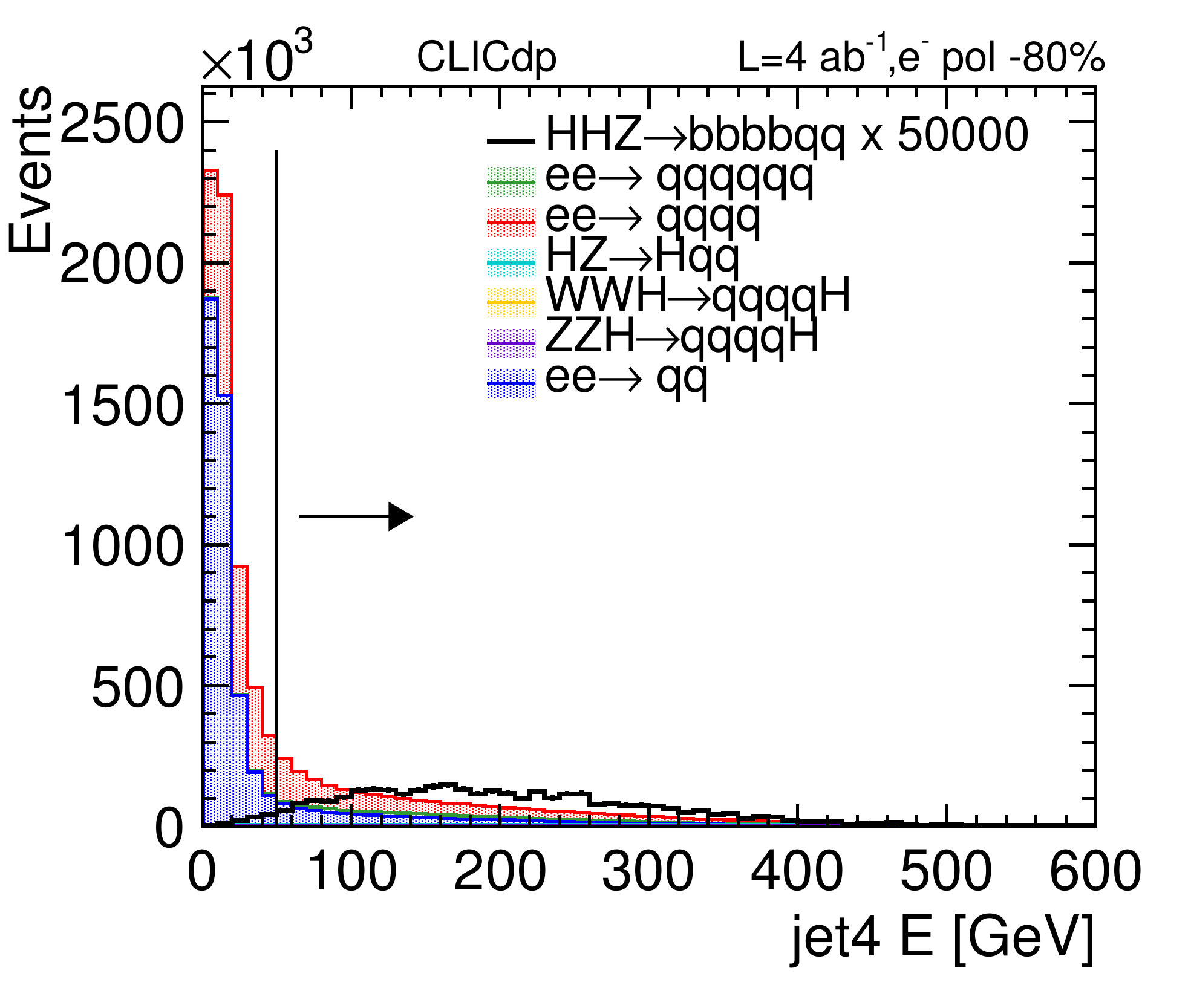}
\end{minipage}
\caption{The energy distribution for the four jets with the largest energies decreasing from left to right for backgrounds events from \zhsm, $\epem\rightarrow\qqbar$, $\epem\rightarrow\qqqq$ , and $\epem\rightarrow \text{qqqqqq}$ combined and signal events $\PZ\PH\PH\rightarrow \qqbar\bb\bb$ weighted by a factor of 50 000. The vertical lines and arrows indicate the signal selection.}
\label{fig:jetenergies}
\end{figure}

In a first attempt boosted decision trees are used to separate signal events and background events using the full available MC statistics in the training. While backgrounds are reduced substantially, the signal cannot be identified with sizeable significance. The MC statistics for backgrounds is limited, and the BDT training is insufficient under these circumstances. In order to facilitate the machine learning processing, different preselections are considered to achieve a larger discriminating power by concentrating on more signal-like background events.
Different preselection cuts are applied, based on BTagging information, mass selections on the reconstruction boson candidates, as well as cuts on the energy and polar angles of jets. The preselection which leads to the best results after tuning of the Boosted Decision Trees is the following:
\begin{itemize}
\item mass selection on the Z candidate (third largest massive boson candidate): \SI{50}{GeV}$<M_{3}<$\SI{150}{GeV}
\item mass selection on the first and second H candidate (order by mass): $M_{1}>\SI{75}{GeV}$, $M_{2}>\SI{75}{GeV}$
\item the BTag sum of the three jets with the largest BTagging values: $\sum \mathrm{BTag}(\mathrm{max\,3})>2.2$
\item polar angles of the leading two jets in energy: $10^{\circ}<\theta\mathrm{(j1)}<170^{\circ}$, $10^{\circ}<\theta\mathrm{(j2)}<170^{\circ}$ 
\item energies of the leading four jets in energy: $E(\mathrm{j1})>\SI{150}{GeV}$, $E(\mathrm{j2})>\SI{100}{GeV}$, $E(\mathrm{j3})>\SI{50}{GeV}$, and $E(\mathrm{j4})>\SI{50}{GeV}$
\end{itemize}
The masses of the boson candidates are shown in Fig.~\ref{fig:bosonMasses} for signal and background events with a signal enhancement of 50 000 in order to emphasise on the shape difference. For the di-quark and four-quark datasets these distributions peak at low values, whereas most of the six-quark dataset appears at mass values similar to the signal events. While for signal events the tail to higher values is significant for the \PH boson candidates (most and second-most massive candidates), the tail to higher mass values is negligible for the \PZ boson candidate. The polar angles for the two jets with the highest energy as well as the BTag sum of the three jets with the largest BTag values are shown in Fig.~\ref{fig:polarangles_BTagSum} for background and signal events with a signal enhancement of 50 000 in order to emphasise on the shape difference. For the signal for both jets the polar angle distributions peak in the central part of the detector, whereas for di-quark, four-quark, and six-quark events the distributions are peaked very forward. For \zhsm events the polar angle distribution of the leading jet is peaked forward as well. Since the signal includes at least four b-quarks, several jets contain B-hadrons in their decay chain, thus the BTag sum distribution of the leading three b-tagged jets peaks at a high value close to 3. Six-quark events from \ttbar contain two b-jets as well, with a peak around 2. For WWH and ZZH events the presence of the \PH boson and its dominant decay into two b-quarks leads to a peak in the BTag sum distribution around 2. While for WWH events the distribution drops at higher values, for ZZH another peak at higher values close to 3 is present due to the fact that the \PZ boson can decay into two b-quarks as well. For di- and four-quark events the distributions peak around 0.5. The energy distributions of the four jets with the largest energy are displayed in Fig.~\ref{fig:jetenergies}. While for signal, but also for the six-quark background the jet energy is typically beyond the cuts of the preselection, for di- and four-quark events the jet energy distributions peak well below the preselection cuts.

As shown in the figures, the preselection is most powerful in rejecting di- and four-quark events. The preselection efficiencies on background and signal datasets are listed in Tab.~\ref{Tab:preselection_efficiency}. About 22\% of signal events are rejected by the preselection, di- and four-boson events are rejected by over 99\%, as well as 96\% of Higgsstrahlung events. Less than 5\% of the six-quark events and less than 4\% of the $\PW\PW\PH$ events remain after the preselection. About 11\% of $\PZ\PZ\PH$ events survive the preselection. Starting from these background events the signal can be extracted with sufficient significance using boosted decision trees.

\begin{table}[hbtp]
 \centering
 \caption{Preselection efficiencies and event numbers\label{Tab:preselection_efficiency} for signal and background events, assuming an integrated luminosity of $\mathrm{L}=\SI{4}{\abinv}$ for runs with negative polarisation P(\Pem)=-80\%, and $\mathrm{L}=\SI{1}{\abinv}$ for runs with positive polarisation P(\Pem)=+80\% for all different final states of the $\Pep\Pem$ collisions:}
 \begin{tabular}{|c|c|c|c|c|c|c|}
\hline
final state & Events & Evts after cut & Efficiency &  Events & Evts after cut & Efficiency  \\
  & -80\%  & -80\%& -80\%, in [\%] &  +80\% & +80\%  & +80\% [in\%]\\
\hline
$\PH\PH\qqbar$, both $\PH\rightarrow\bb$ & 69 & 33 & 48 & 12 & 5.7 & 49 \\
$\PH\PH\qqbar$, all H decays & 167 & 55 & 33 & 29 & 9.5 & 33 \\
$\PH\qqbar$, all \PH & 15300 & 621 & 4.1 & 2670 & 107 & 4.0 \\
$\qqbar$ & 5 070 000 & 3310 & 0.065 & 786 000 & 257 & 0.033 \\
$\qqqq$ & 3 610 000 & 1580 & 0.044 & 120 000 & 143 & 0.12 \\
$\qqqqqq$ & 311 000 & 11000 & 3.5 & 23 900 & 1120 & 4.7 \\
$\PW\PW\PH\rightarrow\qqqq\PH$ & 16 500 & 420 & 2.6 & 518 & 17 & 3.3\\
$\PZ\PZ\PH\rightarrow\qqqq\PH$ & 558 & 61 & 11 & 72 & 8.1 & 11\\
\hline
  \end{tabular}
 \end{table}

\section{Final signal extraction}
\label{BDT}

After the preselection the signal $\PZ\PH\PH\rightarrow \qqbar\bb\bb$ is still a factor of about 500 smaller than the sum of all backgrounds (300 for events with positive electron beam polarisation P(\Pem)=+80\%). After the preselection is applied, the six-quark final state is the dominant background. In order to extract the signal, BDTs are used. Two implementation are considered: BDTs as implemented in the Toolkit for MultiVariate data Analysis (TMVA)~\cite{Hocker:2007ht}, integrated into ROOT~\cite{Antcheva:2009zz}, and the XGBoost (Extreme Gradient Boost) gradient boosting library~\cite{xgboost} as interfaced with scikit-learn. The following variables are used to derive the BDTs:
\begin{itemize}
\item the n-jet resolution thresholds $y_{n-1,n}$ for up to six VLC jets: $y_{12}$, $y_{23}$, $y_{34}$, $y_{45}$, and $y_{56}$, which are all shifted to larger values for signal events, which are more multi-jet like than most background events.
\item the mass values of the three boson candidates $M_{j1}$, $M_{j2}$, and $M_{j3}$ and the angle between the two jets which are combined into both \PH boson candidates $\Delta\alpha_{\PH_{1}}(j1,j2)$, $\Delta\alpha_{\PH_{2}}(j1,j2)$, which are larger for background events.
\item the BTag of the jet with the larger BTag value for each of the boson candidates, as well as the LTag of the jet with the larger light flavour compatibility for each of the boson candidates. The $\mathrm{BTag}_{max}$ distribution peaks closer to 0 for background than for signal events; for the light flavor tag background events show a large peak around 0.8, while for signal a double peak structure appears with a peak close to 0.8 and one around 0. For both \PH candidates the CTag value for the larger c-tagged jet is considered as well. 
\item the sum of the BTag values of the two and the three largest b-tagged jets, which are shifted to larger values for signal; the sum of LTag values of all six jets, which is shifted to considerably larger values for background events.
\item the energy for all six VLC jets, which are higher for signal, and the polar angles for all six VLC jets, which are more central for signal.
\end{itemize}

\begin{figure}[htbp!]
\centering
\begin{minipage}[l]{0.49\textwidth}
\includegraphics[width=1.0\textwidth]{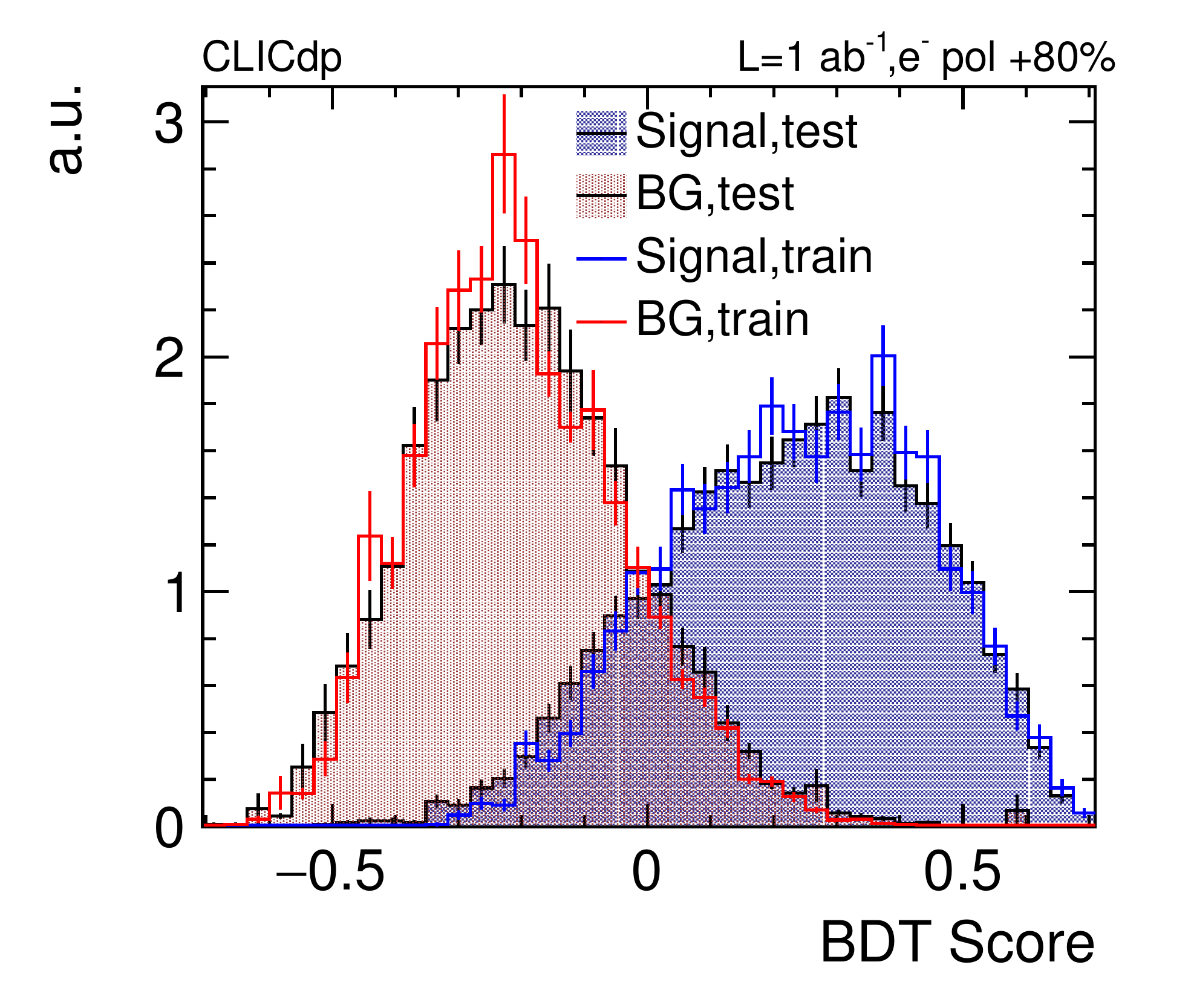}
\end{minipage}
\begin{minipage}[r]{0.49\textwidth}
\includegraphics[width=1.0\textwidth]{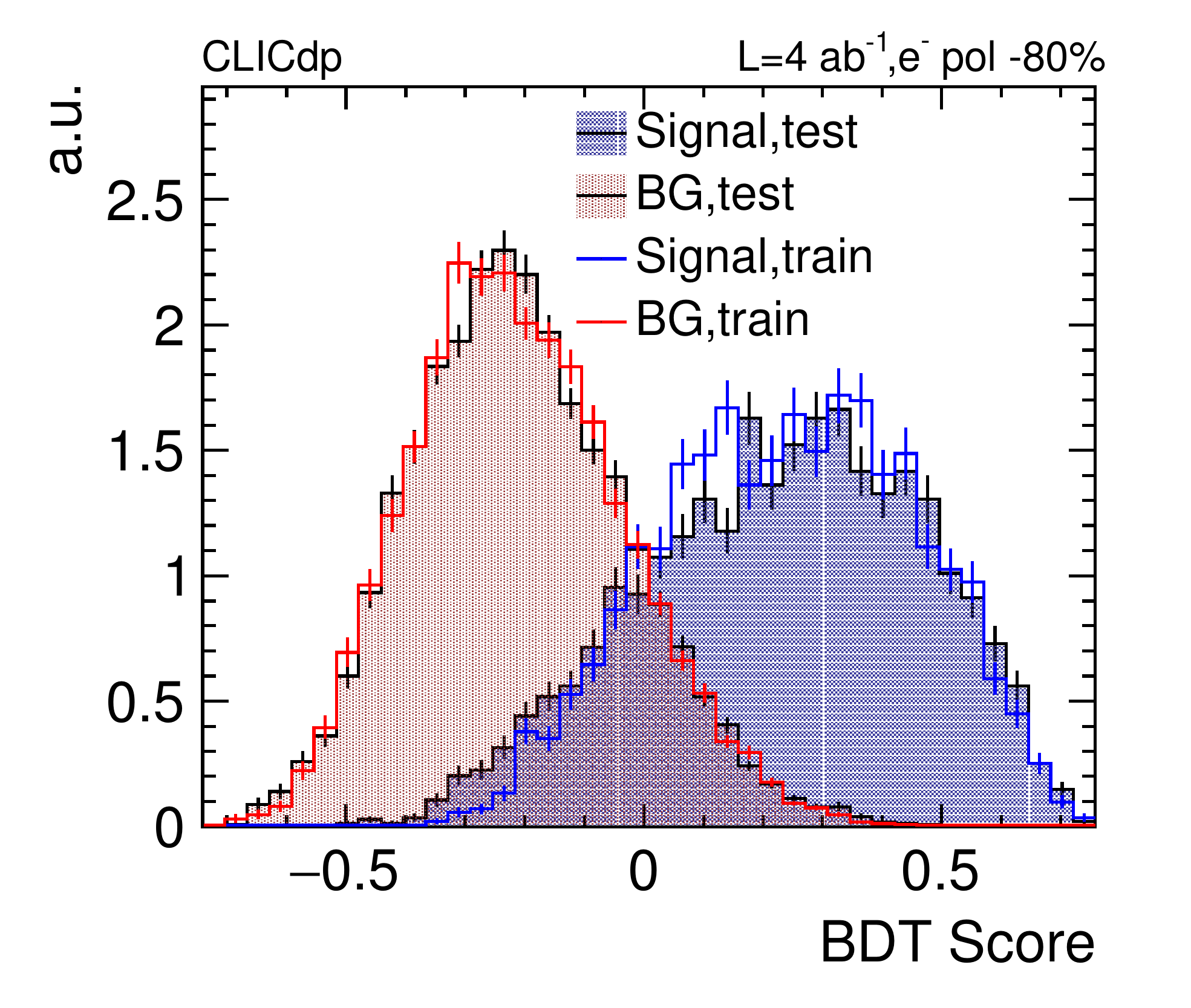}
\end{minipage}
\caption{The distribution of the BDT score for positive (left) and negative (right) electron beam polarisation for the training and testing samples using TMVA.}
\label{fig:BDTscores}
\end{figure}

\begin{figure}[htbp!]
\centering
\begin{minipage}[l]{0.49\textwidth}
\includegraphics[width=1.0\textwidth]{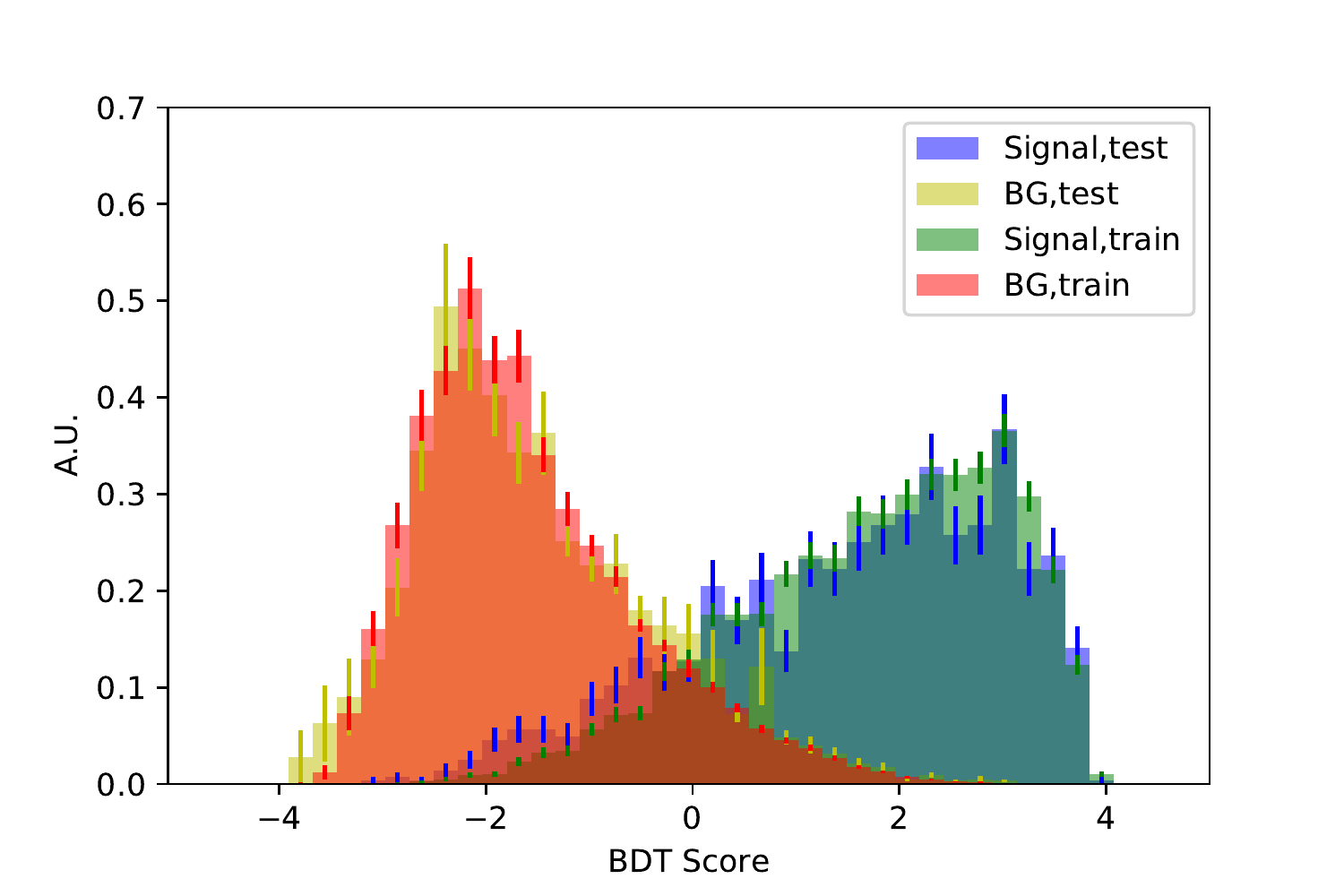}
\end{minipage}
\begin{minipage}[r]{0.49\textwidth}
\includegraphics[width=1.0\textwidth]{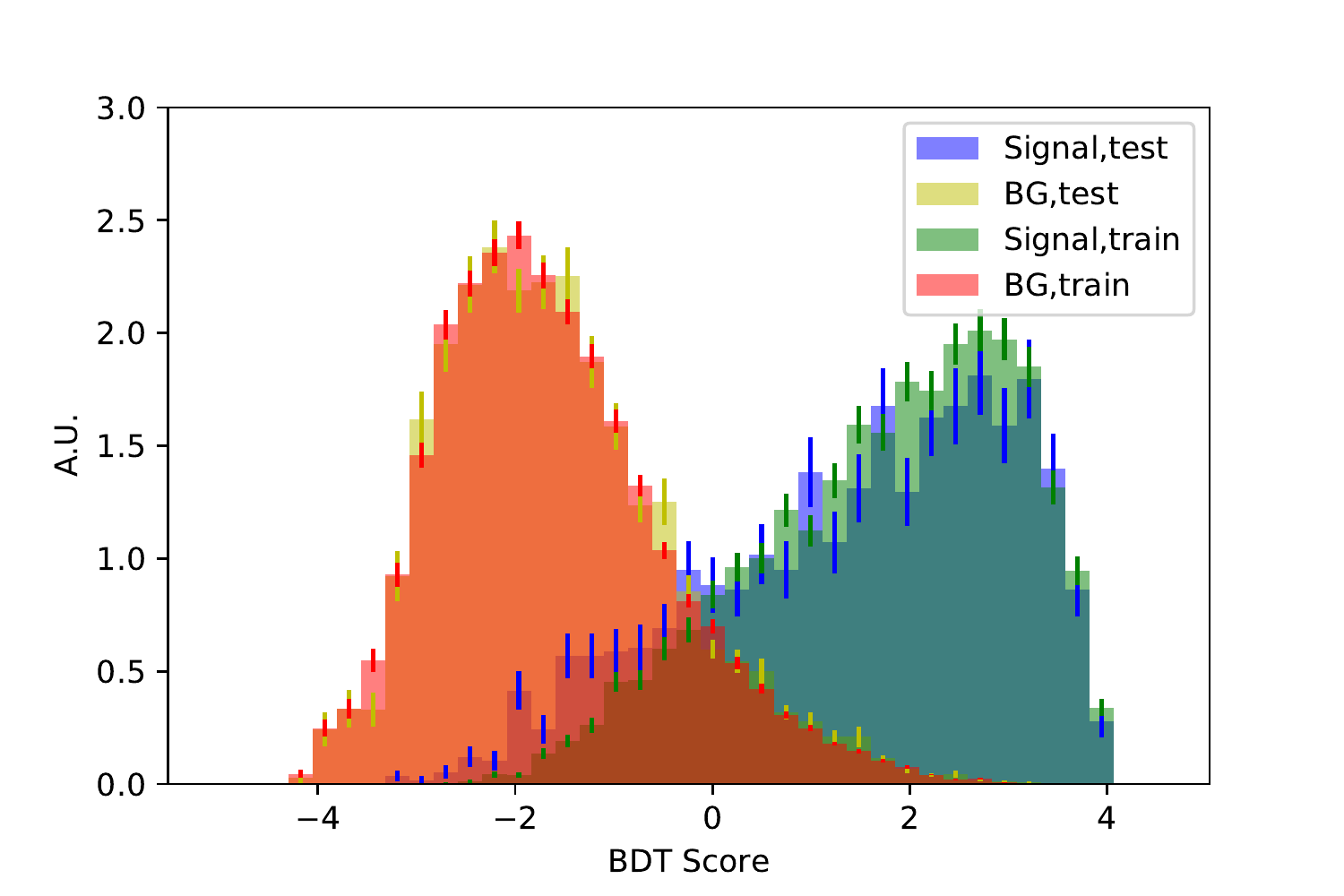}
\end{minipage}
\caption{The distribution of the BDT score for positive (left) and negative (right) electron beam polarisation for the training and testing samples using XGBoost.}
\label{fig:BDTscoresXGBoost}
\end{figure}

For TMVA the best results are achieved using the Gini-index for separation criteria and adaptive boosting instead of gradient boosting. The agreement of the BDT score distributions from the training and the testing datasets for both polarisations in Fig.~\ref{fig:BDTscores} shows that no strong overtraining is observed, despite the limited statistics of the background MC samples. While in TMVA the data is split into testing and training datasets with the ratio 1:1, in XGBoost this ratio is configurable and chosen to be 1:4. In XGBoost we use the gbtree booster and the exact greedy algorithm for split finding. Although a more aggressive boosting was chosen in XGBoost (maximum depth of trees of six compared to four in TMVA) the level of overtraining in XGBoost is on a similar level to that observed for TMVA as Fig.~\ref{fig:BDTscoresXGBoost} illustrates, using the default regularisation values, such as $\lambda=1$ for L2 regularisation.

\section{Results}
\label{sec:summary}

\begin{figure}[htbp!]
\centering
\begin{minipage}[l]{0.49\textwidth}
\includegraphics[width=1.0\textwidth]{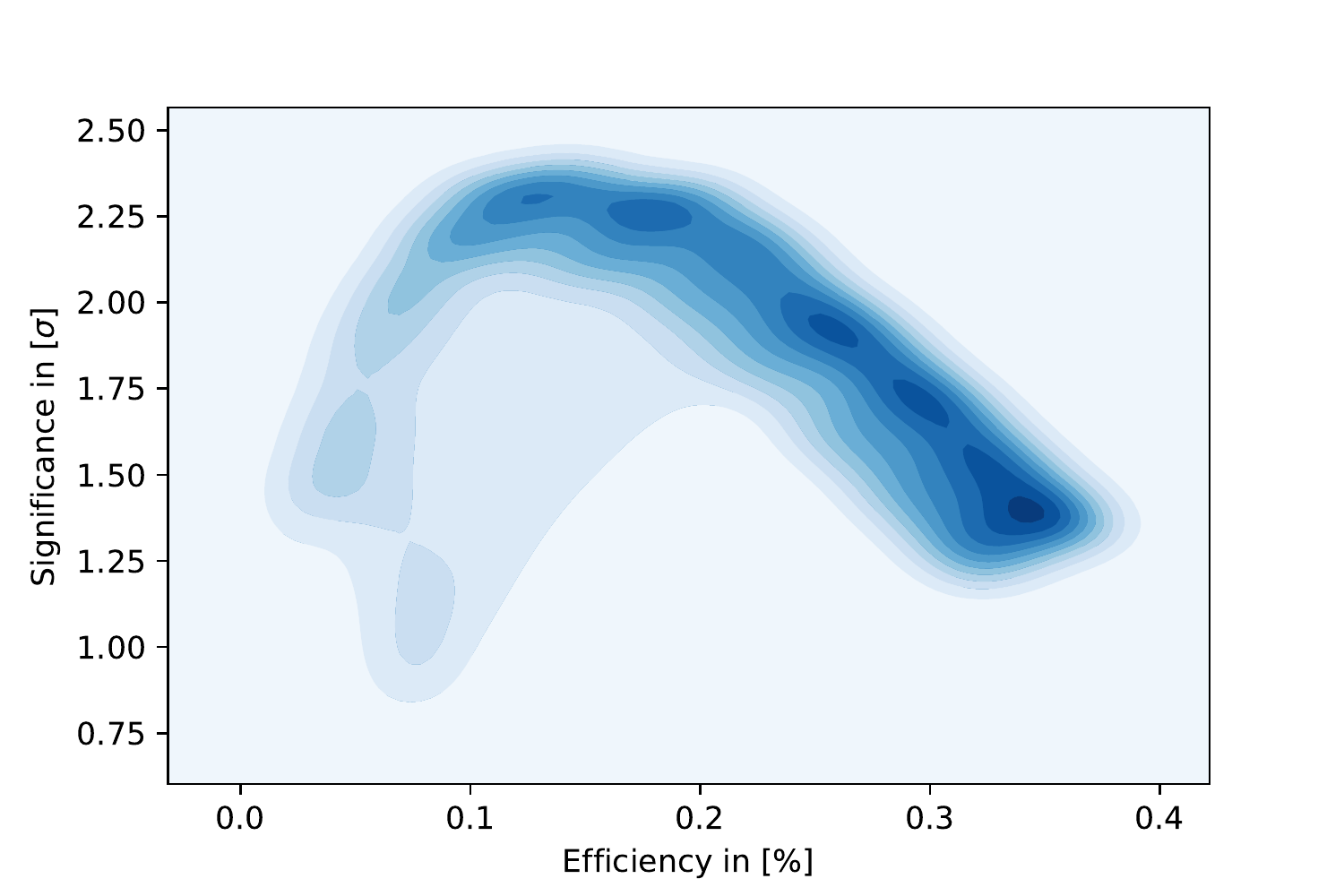}
\end{minipage}
\begin{minipage}[r]{0.49\textwidth}
\includegraphics[width=1.0\textwidth]{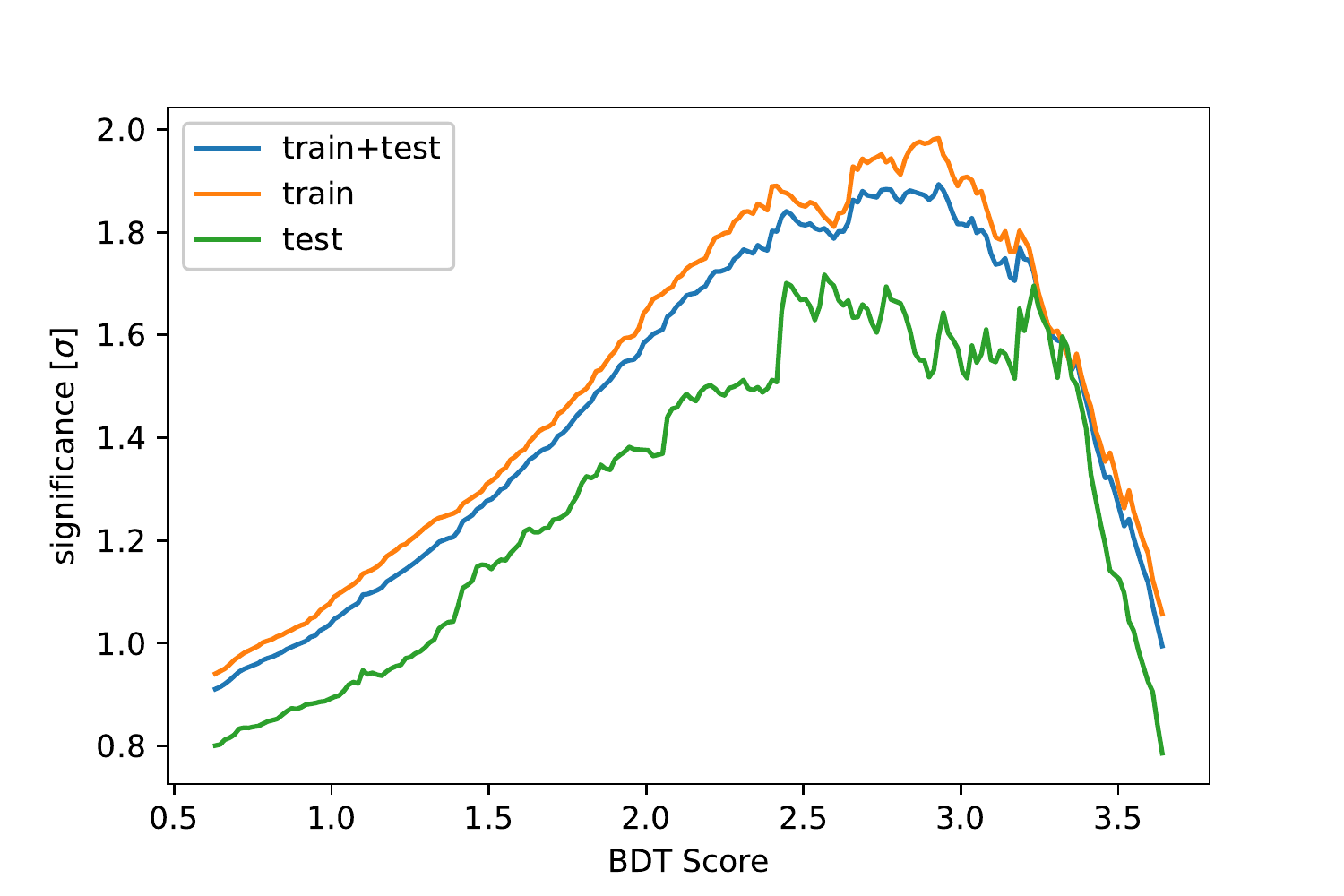}
\end{minipage}
\caption{The significance as function of the signal efficiency (left) and the significance as function of the BDT score using all data, the testing, or only the training data (right) in events with negative electron beam polarisation using XGBoost.}
\label{fig:significance_vs_efficiency}
\end{figure}

The BDTs are trained separately for both polarisation runs, restricting the signal to the most prominent decay channel $\PZ\PH\PH\rightarrow \qqbar\bb\bb$. For the final cross-section result the event numbers of both runs are added up. Using TMVA, the event numbers for both polarisation runs as well as the sum of these numbers are listed in Tab.~\ref{fig:defaulttable}, with a selection on the BDT score of $\mathrm{BDT}>0.385$ for events with negative electron beam polarisation, and $\mathrm{BDT}>0.325$ for events with positive electron beam polarisation. With this selection a significance of about 2.09 $\sigma$ is achieved. Applying the analogous procedure for XGBoost based on the untransformed BDT scores using a maximum tree depth of 6, the combined event numbers are listed in Tab.~\ref{fig:defaulttableXGBoost}, with a selection on the BDT score of $\mathrm{BDT}>2.45$ for events with negative electron beam polarisation, and $\mathrm{BDT}>2.23$ for events with positive electron beam polarisation. With this selection a significance of about 2.38 $\sigma$ is achieved. Using trees with larger depth higher significances can be achieved but at the cost of larger overtraining. The overtraining can be mitigated using more conservative values for other parameters, increasing for example the regularisation parameters $\alpha$, $\lambda$, $\gamma$, or the minimum sum of weights needed to create a child node. Choosing values which keep overtraining on the level observed for depth six or smaller, the achieved significance remains on the level of 2.2-2.4 $\sigma$. Fig.~\ref{fig:significance_vs_efficiency} (left) shows the significance as function of signal efficiency, where the signal is defined as the sum of all $\PZ\PH\PH\rightarrow \qqbar\bb\bb$ events. Significances between 2.1 and 2.3 $\sigma$ are achieved for a relatively broad band of signal efficiencies between 10\% and 22\%.

Considering that the statistical uncertainty of the cross-section determination is more than 40\% the impact of systematic uncertainties on the final result e.g from the flavour tagging shapes, mass and jet scales and resolution is expected to be sub-leading. The significance is studied as function of the BDT score separately for the test and train dataset, scaling both of the event yields to the total event yield. The difference in the significance is typically about 0.2 to 0.25 $\sigma$ (see Fig.~\ref{fig:significance_vs_efficiency} (right)). This 15\% difference can be considered as overtraining systematics, which is sub-leading compared to the statistical uncertainty. BDT tuning on a relaxed preselection with $\sum \mathrm{BTag}(\mathrm{max\,3})>1.0$ achieves a less performant background subtraction. Adding additional input variables in the BDT such as the helicity angles, the larger light flavour tag of the jets of the boson candidates, or the energy ratio between the two jets of each boson candidate lead to slightly worse performances. While this result in the full hadronic channel is not sufficient for a standalone discovery of \zhhsm production at \SI{3}{TeV} CLIC with a significance of slightly above 2~$\sigma$, the sensitivity achieved can contribute to an inclusive double Higgs boson production measurement, and help solving ambiguities in EFT fits using processes with sensitivity to the Higgs self coupling. This is one of the first physics studies using the new detector model and software chain, demonstrating the performance of jet reconstruction and flavour tagging in a complex multi-jet final state at very high energy.

\begin{table}[hbtp]
 \centering
 \caption{Final event numbers\label{fig:defaulttable} for signal and background events using TMVA, assuming an integrated luminosity of $\mathrm{L}=\SI{4}{\abinv}$ for runs with negative polarisation P(\Pem)=-80\%, and $\mathrm{L}=\SI{1}{\abinv}$ for runs with positive polarisation P(\Pem)=+80\% for all different final states of the $\Pep\Pem$ collisions:}
 \begin{tabular}{|c|c|c|c|c|c|c|}
\hline
final state & Events &  Events & Events  \\
  & -80\% & +80\% &  -80\% and +80\% \\
\hline
$\PH\PH\qqbar$, all H decays & $11.11\pm0.22$ & $1.77\pm0.04$ & $13.64\pm0.22$ \\  
$\PH\PH\qqbar$, both $\PH\rightarrow\bb$ & $9.30\pm0.20$ & $1.47\pm0.04$ & $11.37\pm0.21$ \\
$\PH\qqbar$, all \PH & $1.06\pm0.38$ & $0.36\pm0.18$ & $1.43\pm0.42$ \\
$\qqbar$ & $0.00\pm0.00$ & $2.02\pm2.02$& $2.02\pm2.02$ \\
$\qqqq$ & $1.88\pm1.88$ & $0.25\pm0.25$& $2.14\pm1.90$\\
$\qqqqqq$ &  $13.0\pm1.3$ & $2.63\pm0.36$ & $15.6\pm1.4$ \\
$\PW\PW\PH\rightarrow\qqqq\PH$ & $1.57\pm0.52$ & $0.21\pm0.21$& $1.78\pm0.56$\\
$\PZ\PZ\PH\rightarrow\qqqq\PH$ & $4.82\pm0.68$ & $1.01\pm0.16$& $5.83\pm0.70$\\
\hline
  \end{tabular}
 \end{table}
 
\begin{table}[hbtp]
 \centering
 \caption{Final event numbers\label{fig:defaulttableXGBoost} for signal and background events using XGBoost, assuming an integrated luminosity of $\mathrm{L}=\SI{4}{\abinv}$ for runs with negative polarisation P(\Pem)=-80\%, and $\mathrm{L}=\SI{1}{\abinv}$ for runs with positive polarisation P(\Pem)=+80\% for all different final states of the $\Pep\Pem$ collisions:}
 \begin{tabular}{|c|c|c|c|c|c|c|}
\hline
final state & Events &  Events & Events  \\
  & -80\% & +80\% &  -80\% and +80\% \\
\hline
$\PH\PH\qqbar$, all H decays & $13.98\pm0.25$ & $3.43\pm0.05$ & $17.40\pm0.25$ \\
$\PH\PH\qqbar$, both $\PH\rightarrow\bb$  & $11.80\pm0.23$ & $2.74\pm0.05$ & $14.55\pm0.23$ \\
$\PH\qqbar$, all \PH & $1.99\pm0.57$ & $0.64\pm0.24$ & $2.64\pm0.57$ \\
$\qqbar$ & $0.00\pm0.00$ & $0.0\pm0.0$& $0.0\pm0.0$ \\
$\qqqq$ & $1.88\pm1.88$ & $0.25\pm0.25$& $2.14\pm1.90$\\
$\qqqqqq$ &  $17.3\pm1.5$ & $4.33\pm0.45$ & $21.6\pm1.6$ \\
$\PW\PW\PH\rightarrow\qqqq\PH$ & $1.57\pm0.52$ & $0\pm0.0$& $1.57\pm0.52$\\
$\PZ\PZ\PH\rightarrow\qqqq\PH$ & $6.56\pm0.80$ & $1.46\pm0.19$& $8.01\pm0.82$\\
\hline
  \end{tabular}
 \end{table}

\section*{Acknowledgements}
This work benefited from services provided by the ILC Virtual Organisation, supported by the national resource providers of the EGI Federation. This research was done using resources provided by the Open Science Grid, which is supported by the National Science Foundation and the U.S. Department of Energy's Office of Science.
This project has received funding from the European Union's Horizon 2020 Research and Innovation programme under Grant Agreement no. 654168.


\printbibliography[title=References]

@article{CLICdet_note_2017,   
      author =       {Alipour Tehrani, N. and others},  
      title  =        {CLICdet: The post-CDR CLIC detector model},   
      year   =         {2017},   
      month  =        {mar},   
      note   =         {CLICdp-Note-2017-001},   
      url    =          {https://cds.cern.ch/record/2254048} 
}

@article{Sjostrand:2006za,
    author = "Sjostrand, Torbjorn and Mrenna, Stephen and Skands, Peter Z.",
    title = "{PYTHIA 6.4 Physics and Manual}",
    eprint = "hep-ph/0603175",
    archivePrefix = "arXiv",
    reportNumber = "FERMILAB-PUB-06-052-CD-T, LU-TP-06-13",
    doi = "10.1088/1126-6708/2006/05/026",
    journal = "JHEP",
    volume = "05",
    pages = "026",
    year = "2006"
}

@article{xgboost,
    author = {Chen, Tianqi and Guestrin, Carlos},
    year = {2016},
    month = {03},
    eprint = "1603.02754",
    archivePrefix = "arXiv",
    reportNumber = "{KDD '16: Proceedings of the 22nd ACM SIGKDD International Conference on Knowledge Discovery and Data Mining}",
    doi = "10.1145/2939672.2939785",
    title = {XGBoost: A Scalable Tree Boosting System}
}

@article{Kilian:2007gr,
    author = "Kilian, Wolfgang and Ohl, Thorsten and Reuter, Jurgen",
    title = "{WHIZARD: Simulating Multi-Particle Processes at LHC and ILC}",
    eprint = "0708.4233",
    archivePrefix = "arXiv",
    primaryClass = "hep-ph",
    reportNumber = "DESY-11-126, EDINBURGH-2010-36, FR-PHENO-2010-037, SI-HEP-2010-18",
    doi = "10.1140/epjc/s10052-011-1742-y",
    journal = "Eur. Phys. J. C",
    volume = "71",
    pages = "1742",
    year = "2011"
}

@article{Leogrande:2019dzm,
    author = "Leogrande, Emilia and Roloff, Philipp and Schnoor, Ulrike and Weber, Matthias",
    archivePrefix = "arXiv",
    eprint = "1911.02523",
    month = "11",
    primaryClass = "hep-ex",
    reportNumber = "CLICdp-Note-2019-005",
    title = "{All-hadronic HZ production at high energy at 3 TeV CLIC}",
    year = "2019"
}

@article{Hocker:2007ht,
	author 	=	"Hoecker, Andreas and Speckmayer, Peter and Stelzer, Joerg and Therhaag, Jan and von Toerne, Eckhard and Voss, Helge",
	title 	 = 	"{TMVA: Toolkit for Multivariate Data Analysis}",
	journal 	 = 	"PoS",
	volume 	 = 	"ACAT",
	year 	 = 	"2007",
	pages 	 = 	"040",
	eprint 	 = 	"physics/0703039",
	archivePrefix 	 = 	"arXiv",
	SLACcitation 	 = 	"%%CITATION = PHYSICS/0703039;%%"
}

@article{Roloff:2019crr,
      author         = "Roloff, Philipp and Schnoor, Ulrike and Simoniello, Rosa
                        and Xu, Boruo",
      title          = "{Double Higgs boson production and Higgs self-coupling
                        extraction at CLIC}",
      collaboration  = "CLICdp",
      year           = "2019",
      eprint         = "1901.05897",
      archivePrefix  = "arXiv",
      primaryClass   = "hep-ex",
      reportNumber   = "v1: CLICdp-Note-2018-006, v2: CLICdp-Pub-2019-007,
                        CLICdp-Note-2018-006",
      SLACcitation   = "%%CITATION = ARXIV:1901.05897;%%"
}

@article{Boronat:2016tgd,
      author         = "Boronat, M. and Fuster, J. and Garcia, I. and Roloff, Ph.
                        and Simoniello, R. and Vos, M.",
      title          = "{Jet reconstruction at high-energy electron-positron
                        colliders}",
      journal        = "Eur. Phys. J.",
      volume         = "C78",
      year           = "2018",
      number         = "2",
      pages          = "144",
      doi            = "10.1140/epjc/s10052-018-5594-6",
      eprint         = "1607.05039",
      archivePrefix  = "arXiv",
      primaryClass   = "hep-ex",
      reportNumber   = "CLICDP-PUB-2017-002",
      SLACcitation   = "%%CITATION = ARXIV:1607.05039;%%"
}

@article{Antcheva:2009zz,
      author         = "Antcheva, I. and others",
      title          = "{ROOT: A C++ framework for petabyte data storage,
                        statistical analysis and visualization}",
      journal        = "Comput. Phys. Commun.",
      volume         = "180",
      year           = "2009",
      pages          = "2499-2512",
      doi            = "10.1016/j.cpc.2009.08.005",
      eprint         = "1508.07749",
      archivePrefix  = "arXiv",
      primaryClass   = "physics.data-an",
      reportNumber   = "FERMILAB-PUB-09-661-CD",
      SLACcitation   = "%%CITATION = ARXIV:1508.07749;%%"
}

@article{Arominski:2018uuz,
      author         = {Arominski, Dominik and others},
      title          = {A detector for CLIC: main parameters and performance},
      collaboration  = {CLICdp Collaboration},
      year           = {2018},
      eprint         = {1812.07337},
      archivePrefix  = {arXiv},
      primaryClass   = {physics.ins-det},
      reportNumber   = {CLICdp-Note-2018-005},
      SLACcitation   = {%%CITATION = ARXIV:1812.07337;%%}
}

@ARTICLE{Tran:2017tgrSoftwareCompensation,
  author = {Tran, H.L. and Kr{\"u}ger, K. and Sefkow, F. and Green, S. and Marshall,
	J.S and Thomson,M.A. and Simon, F.},
  title = {Software compensation in Particle Flow reconstruction},
  journal = {Eur. Phys. J.},
  year = {2017},
  volume = {C77},
  pages = {698},
  number = {10},
  archiveprefix = {arXiv},
  doi = {10.1140/epjc/s10052-017-5298-3},
  eprint = {1705.10363},
  primaryclass = {physics.ins-det},
  reportnumber = {DESY-17-083, MPP-2017-98},
  slaccitation = {%%CITATION = ARXIV:1705.10363;%%}
}

@article{deBlas:2018mhx,
    author = "Franceschini, R. and others",
    title = "{The CLIC Potential for New Physics}",
    eprint = "1812.02093",
    archivePrefix = "arXiv",
    primaryClass = "hep-ph",
    reportNumber = "CERN-TH-2018-267, CERN-2018-009-M",
    doi = "10.23731/CYRM-2018-003",
    month = "12",
    year = "2018"
}

@article{Aicheler:2019dhf,
      author         = "Aicheler, M. and Burrows, P. N. and Catalan Lasheras, N.
                        and Corsini, R. and Draper, M. and Osborne, J. and
                        Schulte, D. and Stapnes, S. and Stuart, M. J.",
      title          = "{The Compact Linear Collider (CLIC) - Project
                        Implementation Plan}",
      collaboration  = "CLIC Collaboration",
      doi            = "10.23731/CYRM-2018-004",
      year           = "2019",
      eprint         = "1903.08655",
      archivePrefix  = "arXiv",
      primaryClass   = "physics.acc-ph",
      reportNumber   = "CERN-2018-010-M",
      SLACcitation   = "%%CITATION = ARXIV:1903.08655;%%"
}

@article{Abramowicz:2016zbo,
      author         = "Abramowicz, H. and others",
      title          = {Higgs Physics at the CLIC electron-positron linear collider},
      journal        = {Eur. Phys. J.},
      volume         = {C77},
      pages          = {475},
      doi            = {10.1140/epjc/s10052-017-4968-5},
      year           = {2017},
      eprint         = "1608.07538",
      archivePrefix  = "arXiv",
      primaryClass   = "hep-ex",
      reportNumber   = "CLICDP-PUB-2016-001",
      SLACcitation   = "%%CITATION = ARXIV:1608.07538;%%"
}

@article{CLIC:2016zwp,
      author         = "Boland, M J and others",
      editor         = "Lebrun, P and Linssen, L and Schulte, D and Sicking, E
                        and Stapnes, S and Thomson, M A and Burrows, P N",
      title          = "{Updated baseline for a staged Compact Linear Collider}",
      collaboration  = "CLIC and CLICdp Collaborations",
      doi            = "10.5170/CERN-2016-004",
      year           = "2016",
      eprint         = "1608.07537",
      archivePrefix  = "arXiv",
      primaryClass   = "physics.acc-ph",
      reportNumber   = "CERN-2016-004",
      SLACcitation   = "%%CITATION = ARXIV:1608.07537;%%"
}

@article{Suehara:2015ura,
      author         = "Suehara, Taikan and Tanabe, Tomohiko",
      title          = "{LCFIPlus: A Framework for Jet Analysis in Linear
                        Collider Studies}",
      journal        = "Nucl. Instrum. Meth.",
      volume         = "A808",
      year           = "2016",
      pages          = "109-116",
      doi            = "10.1016/j.nima.2015.11.054",
      eprint         = "1506.08371",
      archivePrefix  = "arXiv",
      primaryClass   = "physics.ins-det",
      SLACcitation   = "%%CITATION = ARXIV:1506.08371;%%"
}

@article{Brondolin:2019awm,
    author = {Brondolin, Erica and Leogrande, Emilia and Hynds, Daniel and Gaede, Frank and Petric, Marko and Sailer, Andre and Simoniello, Rosa},
    collaboration = "CLICdp",
    title = "{Conformal tracking for all-silicon trackers at future electron--positron colliders}",
    eprint = "1908.00256",
    archivePrefix = "arXiv",
    primaryClass = "physics.ins-det",
    reportNumber = "CLICdp-Pub-2019-003",
    doi = "10.1016/j.nima.2019.163304",
    journal = "Nucl. Instrum. Meth. A",
    volume = "956",
    pages = "163304",
    year = "2020"
}

@article{Frank:2015ivo,
      author         = "Frank, M. and Gaede, F. and Nikiforou, N. and Petric, M.
                        and Sailer, A.",
      title          = "{DDG4 A Simulation Framework based on the DD4hep Detector
                        Description Toolkit}",
      booktitle      = "{Proceedings, 21st International Conference on Computing
                        in High Energy and Nuclear Physics (CHEP 2015): Okinawa,
                        Japan, April 13-17, 2015}",
      journal        = "J. Phys. Conf. Ser.",
      volume         = "664",
      year           = "2015",
      number         = "7",
      pages          = "072017",
      doi            = "10.1088/1742-6596/664/7/072017",
      SLACcitation   = "%%CITATION = 00462,664,072017;%%"
}

@article{Sailer:2017rnh,
      author         = "Sailer, A. and Frank, M. and Gaede, F. and Hynds, D. and
                        Lu, S. and Nikiforou, N. and Petric, M. and Simoniello, R.
                        and Voutsinas, G.",
      title          = "{DD4Hep based event reconstruction}",
      booktitle      = "{Proceedings, 22nd International Conference on Computing
                        in High Energy and Nuclear Physics (CHEP2016): San
                        Francisco, CA, October 14-16, 2016}",
      collaboration  = "CLICdp and ILD Collaborations",
      journal        = "J. Phys. Conf. Ser.",
      volume         = "898",
      year           = "2017",
      number         = "4",
      pages          = "042017",
      doi            = "10.1088/1742-6596/898/4/042017",
      reportNumber   = "CLICdp-Conf-2017-002",
      SLACcitation   = "%%CITATION = 00462,898,042017;%%"
}

@article{Agostinelli:2002hh,
      author         = "Agostinelli, S. and others",
      title          = "{GEANT4: A Simulation toolkit}",
      collaboration  = "GEANT4 Collaboration",
      journal        = "Nucl. Instrum. Meth.",
      volume         = "A506",
      year           = "2003",
      pages          = "250-303",
      doi            = "10.1016/S0168-9002(03)01368-8",
      reportNumber   = "SLAC-PUB-9350, FERMILAB-PUB-03-339",
      SLACcitation   = "%%CITATION = NUIMA,A506,250;%%"
}

@article{Cacciari:2011ma,
      author         = {Cacciari, Matteo and Salam, Gavin P. and Soyez, Gregory},
      title          = {FastJet User Manual},
      journal        = {Eur. Phys. J.},
      volume         = {C72},
      year           = {2012},
      pages          = {1896},
      doi            = {10.1140/epjc/s10052-012-1896-2},
      eprint         = {1111.6097},
      archivePrefix  = {arXiv},
      primaryClass   = {hep-ph},
      reportNumber   = {CERN-PH-TH-2011-297},
      SLACcitation   = {%%CITATION = ARXIV:1111.6097;%%}
}

@article{Schulte:382453,
      author        = "Schulte, Daniel",
      title         = "{Beam-Beam Simulations with GUINEA-PIG}",
      month         = "Mar",
      year          = "1999",
      reportNumber  = "CERN-PS-99-014-LP",
      url           = "https://cds.cern.ch/record/382453",
}

@article{Robson:2018zje,
      author         = "Robson, Aidan and Roloff, Philipp",
      title          = "{Updated CLIC luminosity staging baseline and Higgs
                        coupling prospects}",
      year           = "2018",
      eprint         = "1812.01644",
      archivePrefix  = "arXiv",
      primaryClass   = "hep-ex",
      reportNumber   = "CLICdp-Note-2018-002",
      SLACcitation   = "%%CITATION = ARXIV:1812.01644;%%"
}

@article{Marshall:2015rfa,
      author         = {Marshall, J. S. and Thomson, M. A.},
      title          = {The Pandora Software Development Kit for Pattern
                        Recognition},
      journal        = {Eur. Phys. J.},
      volume         = {C75},
      year           = {2015},
      number         = {9},
      pages          = {439},
      doi            = {10.1140/epjc/s10052-015-3659-3},
      eprint         = {1506.05348},
      archivePrefix  = {arXiv},
      primaryClass   = {physics.data-an},
      SLACcitation   = {%%CITATION = ARXIV:1506.05348;%%}
}

@article{Marshall:2012ry,
      author         = {Marshall, J. S. and M{\"u}nnich, A. and Thomson, M. A.},
      journal        = {Nucl. Instrum. Meth.},
      title = {Performance of Particle Flow Calorimetry at CLIC},
      volume         = {A700},
      year           = {2013},
      pages          = {153-162},
      doi            = {10.1016/j.nima.2012.10.038},
      eprint         = {1209.4039},
      archivePrefix  = {arXiv},
      primaryClass   = {physics.ins-det},
      reportNumber   = {CU-HEP-12-12, AIDA-PUB-2013-002},
      SLACcitation   = {%%CITATION = ARXIV:1209.4039;%%}
}
\

\end{document}